\documentclass[superscriptaddress, amsmath,amssymb, aps,prb,floatfix,twocolumn]{revtex4-1}
\usepackage[utf8]{inputenc}
\usepackage{color}

\usepackage[usenames,dvipsnames,table]{xcolor}
\definecolor{AppRed}{RGB}{255,4,4}
\definecolor{DarkRed}{RGB}{175,4,4}
\definecolor{AppGreen}{RGB}{0, 128, 0}%{40,50,100}
\definecolor{AppBlue}{RGB}{23,23,255}
\definecolor{AppOrange}{RGB}{255,165,0}
\usepackage{svg}
\usepackage{amsmath}

\usepackage[verbose]{newunicodechar}
\newunicodechar{−}{-}
\newunicodechar{ħ}{$\hbar$}
\newunicodechar{ω}{$\omega$}

\usepackage{graphicx}% Include figure files
\usepackage{xspace}
\usepackage{xfrac}
\usepackage{placeins}
\usepackage{subcaption}
\usepackage{bm}% bold math
\usepackage[hidelinks]{hyperref}% add hypertext capabilities
\graphicspath{{Figures/}}
\newcommand{\ymo}{$h$-YMnO$_3$\xspace}
\renewcommand{\exp}[1]{\mathrm{exp}\left(#1\right)}

\begin{document}
%\title{Characterizing the diffuse continuum excitations in \texorpdfstring{\textit{h}-YMnO$_3$}{h-YMnO3}}
\title{Characterizing the diffuse continuum excitations in the classical spin liquid \texorpdfstring{\textit{h}-YMnO$_3$}{h-YMnO3}}
\author{Jakob Lass}
\email{jakob.lass@psi.ch}
\affiliation{Nanoscience Center, Niels Bohr Institute, University of Copenhagen, 2100 Copenhagen \O, Denmark}
\affiliation{Laboratory for Neutron Scattering, Paul Scherrer Institute, 5232 Villigen, Switzerland}
\affiliation{Laboratory for Quantum Magnetism, Ecole Polytechnique Fédérale de Lausanne (EPFL), CH-1015 Lausanne, Switzerland}
\author{Emma Y. Lenander}
\affiliation{Nanoscience Center, Niels Bohr Institute, University of Copenhagen, 2100 Copenhagen \O, Denmark}
\author{Kristine M. L. Krighaar}
\affiliation{Nanoscience Center, Niels Bohr Institute, University of Copenhagen, 2100 Copenhagen \O, Denmark}
\author{Tara N. To\v{s}i\'c} % Niamh
\affiliation{Materials Theory, ETH Zürich, Wolfgang-Pauli-Strasse 27, 8093 Zurich, Switzerland}
\author{Dharmalingam Prabhakaran}
\affiliation{Clarendon Laboratory, Department of Physics, University of Oxford, Oxford OX1 3PU, United Kingdom}
\author{Pascale P. Deen}
\affiliation{European Spallation Source ESS ERIC, Box 176, SE-221 00 Lund, Sweden}
\affiliation{Nanoscience Center, Niels Bohr Institute, University of Copenhagen, 2100 Copenhagen \O, Denmark}
\author{Sofie Holm-Janas}
\affiliation{Department of Physics, Technical University of Denmark, 2800 Kgs.~Lyngby, Denmark}
\affiliation{Nanoscience Center, Niels Bohr Institute, University of Copenhagen, 2100 Copenhagen \O, Denmark}
\author{Kim Lefmann}
\email{lefmann@nbi.ku.dk }
\affiliation{Nanoscience Center, Niels Bohr Institute, University of Copenhagen, 2100 Copenhagen \O, Denmark}
\date{\today}

\begin{abstract}
We extend previous inelastic neutron scattering results on the geometrically frustrated antiferromagnet hexagonal-YMnO$_3$, which has been suggested to belong to the class of classical spin liquids. We extend the energy transfer coverage of the diffuse signal up to 6.9~meV within a wide temperature range around the ordering temperature,  $T_{\rm N}$. The two distinct diffuse signals in the $a$-$b$ plane, the signal localized at $\Gamma^\prime$ and the scattering intensity connecting $\Gamma^\prime$ points over the $\mathrm{M}^\prime$, are shown to be only weakly energy dependent. In addition, an external magnetic field of up to 10.5~T applied along $c$ is shown to have no effect on the diffuse signal. In the orthogonal scattering plane, the signals are shown to be dependent on $l$ only through the magnetic form factor, showing that the correlations are purely two-dimensional, and supporting its origin to be the frustrated Mn$^{3+}$ triangles. This result is corroborated by atomistic spin dynamics simulations showing similar scattering vector and temperature behaviours. Lastly, data for the spin wave scattering in the ($h$, 0, $l$) plane allow for a discussion of the magnetic ground state where better agreement is found between the data and an ordered structure of the $\Gamma_1$ or $\Gamma_3$ symmetry, albeit crystal electric field arguments dismisses the $\Gamma_1$ as possibility.

\end{abstract}
\maketitle

\section{Introduction}
The search for novel phases of materials with intriguing properties is one of the main quests in material science.
One major branch concerns geometrically frustrated magnets, in which competing interactions suppress the magnetic ordering.\cite{Balents2010} Within this research field, quantum spin liquids (QSLs) stand out as a widely sought-after phase, characterized as a fluid-like state where spins exhibit strong correlations and show persistent fluctuation even at temperatures of absolute zero.\cite{Savary2017, Broholm2020, Wen2019} A defining theoretical feature of QSLs is their long-range entanglement, but its experimental verification is complicated by {\em e.g.}\ the presence of other effects, such as impurities, lattice distortions, and non-perfect realizations of the theoretical systems.\cite{Broholm2020, Chamorro2021, Knolle2019} Instead, the hunt continues for frustrated spin-1/2 systems that show no magnetic order at the lowest reachable temperatures, while displaying diffuse and broad excitation continua. These features are taken as the fingerprint of QSL behaviour.\cite{Wen2019, Knolle2019} 

However, one may take other, more pragmatic routes to study spin liquids. Lessening the requirements of a low quantum spin number, {\em i.e.}\ extending to the classical spin limit, still allows for high levels of frustration. In general, these systems do order magnetically, but their ordering temperature, $T_{\rm N}$, is significantly reduced as compared to the Curie-Weiss temperature, $T_{\rm CW}$, obtained from high temperature susceptibility measurements. In such systems, long-range entanglement is impossible, but the systems still have many features very similar to supposed QSLs; in particular a diffuse continuum of excitations near the magnetic ordering. These systems have loosely been referred to as ``cooperative paramagnets'' or ``spin-liquid-like'' systems, while perhaps a more suitable name would be Classical Spin Liquids (CSLs). Unfortunately, a rigorous definition of this material class is still lacking, but one possible defining property has been proposed by some of us to be the presence of diffuse, correlated continuum excitations in the paramagnetic phase. \cite{Janas2021}

\begin{figure}[tbh] % Generated by C:\Users\lass_j\Documents\YMnO3\GenerateMagneticGroundStateFigures.m, 09/11-23 JL
    \iffalse
    \begin{subfigure}[b]{0.24\textwidth}
         \centering
         \includegraphics[trim={1.8cm 0 2.1cm 2.0cm},clip,width=\textwidth]{Figures/Structures/G1.png}
         \caption{$\Gamma_1$}%$\Delta E$ = 0.4 meV}
     \end{subfigure}%
     \begin{subfigure}[b]{0.24\textwidth}
         \centering
         \includegraphics[trim={1.8cm 0 2.1cm 2.0cm},clip,width=\textwidth]{Figures/Structures/G2.png}
         \caption{$\Gamma_2$}
     \end{subfigure}
     \begin{subfigure}[b]{0.24\textwidth}
         \centering
         \includegraphics[trim={1.8cm 0 2.1cm 2.0cm},clip,width=\textwidth]{Figures/Structures/G3.png}
         \caption{$\Gamma_3$}
     \end{subfigure}%
     \begin{subfigure}[b]{0.24\textwidth}
         \centering
         \includegraphics[trim={1.8cm 0 2.1cm 2.0cm},clip,width=\textwidth]{Figures/Structures/G4.png}
         \caption{$\Gamma_4$}
     \end{subfigure}
    \fi
    \includesvg[inkscapelatex=false]{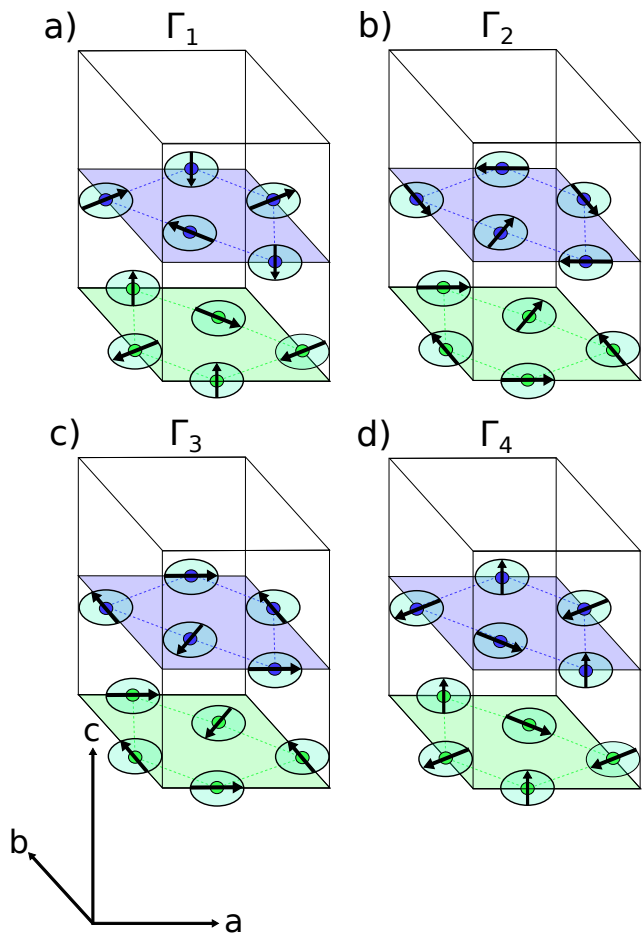}
    \caption{The four allowed magnetic ground state representations, $\Gamma_1$ ($P6_3cm$), $\Gamma_2$ ($P6_3c^\prime m^\prime$), $\Gamma_3$ ($P6^\prime_3cm^\prime$) and $\Gamma_4$ ($P6^\prime_3c^\prime m$), for \ymo based on symmetry analysis\cite{Holm2018} and plotted using SpinW.\cite{Toth15} Only magnetic moments are shown, where green moments are located at the lattice plane $z$ = 0 and blue at $z$ = $\frac12$.}
    \label{fig:GroundStates}
\end{figure}

Hexagonal yttrium manganese (\ymo ) is one candidate system within this proposed class of CSLs. It is a frustrated magnet featuring a stacking of triangular layers of antiferromagnetically coupled  Mn$^{3+}$ spin 2 along the crystallographic $c$ axis. We note that \ymo is also extensively studied for its multiferroic properties.\cite{VanAken2004} Magnetic long range order sets in at $T_{\rm N}=71$~K \cite{Roessli2005,Holm-Dahlin2018} where a two-dimensional 120$^\circ$ spin structure develops within the triangular layers. The exact stacking and spin orientation have been discussed by many authors.\cite{Holm2018, Howard2013, Singh2013, Frohlich1998, Munoz2000, Fiebig2000, Brown2006} Recent in-field spin wave measurements limited the magnetic structure to be described by either the $\Gamma_1$ or $\Gamma_3$ symmetry group, shown in Fig.~\ref{fig:GroundStates}, corresponding to the $P6_3cm$ and $P6_3^{\prime}c^{\prime}m$ symmetries, respectively. \cite{Holm2018} Brown et al., however, suggested that the ordering falls within the $P6^\prime_3$ symmetry with a tilting angle $\psi$ of 11$^\circ$ out of the $(a,b)$-plane. \cite{Brown2006}

The spin wave spectrum measured by inelastic neutron scattering in Ref.~\onlinecite{Holm2018} is well described by a strong isotropic antiferromagnetic nearest-neighbour interaction of $J=2.4$ meV within the triangles with a small planar anisotropy of $D\approx0.312$ meV and an out-of-plane coupling $J_z\approx0.144$ meV. Despite the measured sample containing 2 \% Eu doping, the overall interaction ratios are still representative for the undoped compound as the behaviour of the pure and 2\% doped systems is very similar.\cite{Holm-Dahlin2018,Holm-Dahlin2018Erratum} Due to the multiferroic nature of \ymo, an additional coupling exists between the magnetic and lattice degrees of freedom, resulting in avoided crossings between magnon and phonon dispersions, not further discussed in this paper.\cite{Holm2018,Petit2008} The frustration is manifested in a large discrepancy between the magnetic ordering temperature and the Curie-Weiss temperature, $\theta_{\rm CW}=-500$ K,\cite{Roessli2005,Lee2008} giving a frustration ratio of $f=|\theta_{\rm CW}|/T_{\rm N} = 6.9$.

In a recent publication,\cite{Janas2021} some of us investigated the diffuse continuum excitation in \ymo and linked the behaviour to the definition of classical spin liquids. This investigation was performed using neutron spectroscopy measurements of excitations within the ($a,b$) crystallographic plane where magnetic, gapless excitations were shown to be present both far above the magnetic ordering temperature (up to $0.9~T_{CW}$) as well as far into the ordered phase. That is, the diffuse signals remained well-defined even at 4~K ($0.06~T_N$), where they co-existed with the spin waves. The correlation length of the excitations increased with decreasing temperature, while remaining finite even at 4~K. The diffuse excitations were speculated to be two-dimensional spin correlations of the Mn$^{3+}$ spin clusters within the triangular layers, and were modelled with critical spin-spin correlations.\cite{Janas2021}\\

In the present work, we extend the inelastic measurements and thereby advance the understanding of the diffuse excitations in \ymo . We verify our previous speculation that the spin fluctiations are purely two-dimensional, in agreement with new atomistic spin dynamic simulations of \ymo. Furthermore, we show that they persist to large excitation energies, up to at least 7~meV with some broadening, which largely follows that of the low-temperature spin wave signal. The same result was seen in the simulations,\cite{Tosic2023} further explained below. 

One aspect not touched upon in previous literature is the possibility of the diffuse magnetic scattering being dependent on external magnetic fields. Specifically, an applied field along the crystallographic $c$ direction could  serve to induce a spin canting with a potential impact on the spin correlations. In contrast to the spin waves, the diffuse excitations are found to be independent of an applied magnetic field up to 10.5~T. 

By performing inelastic neutron scattering studies in the ($h$, 0, $l$) plane we find a weak, not previously encountered signal which we speculate to be magnetic. This would advocate for the magnetic ordering of the $\Gamma_1$ type as to match the magnon periodicity. However, according to the crystal field environment, this symmetry is disallowed. In section \ref{sec:spinwave} we discuss the implication of our results on the most likely magnetic symmetry group of the system.%\cite{Das2014}

\section{Experimental Methods}
We have performed inelastic neutron scattering studies on \ymo, where we directly probe the magnetic dynamic structure factor $S(\mathbf{q},\hbar\omega)$ as a function of scattering wavevector $\mathbf{q}$ and energy transfer $\hbar \omega$. Both experiments were performed on a single crystal of pure \ymo, also used in Ref.~\onlinecite{Janas2021}.%, which has a mass of 3.2~g, reduced from the previous 3.4~g used in the ($h$,$k$,0) setup due to small crystallites detaching during transportation and realignment.  

An overall picture of the excitations were obtained through experiments performed using the cold-neutron spectrometer CAMEA at the neutron source SINQ at the Paul Scherrer Institut, Switzerland.\cite{Groitl2016,Lass2023} We were able to obtain quasi-continuous coverage in the three dimensional space spanned by two reciprocal lattice vectors and the energy axis by utilizing its multiplexing analyzers together with the prismatic analyzer concept, \cite{Birk2014} as well as the large detector array. All data were converted and treated using the dedicated software package MJOLNIR, version 1.3.1.  \cite{Lass2020MJOLNIR,jakob_lass_2023_8183140}

In the first setup, the crystal was aligned in the ($h$,$k$,0) scattering plane and placed inside a 11~T split-coil cryomagnet, MB11, from Oxford Instruments. Data were taken at two temperatures, 60~K and 105~K, in both zero field and in an applied field of 10.5~T. Data were acquired using CAMEA by performing 180$^\circ$ sample rotation scans in 1$^\circ$ steps for 4 different incoming energies, (5.25, 6.85, 8.45, and 10.05~meV) with full scattering angle coverage for the lower 3 energies and low $Q$ coverage for the last, covering excitations from 0.05~meV up to 6.9~meV and using a monitor of 100\,000 corresponding to $\sim$23 s per step. The coverage of the measured data at an energy transfer of $\hbar\omega = 1.0 \pm 0.1$~meV is shown in Fig.~\ref{fig:Qmap_BZ_p1}~(a), along with the nomenclature for high symmetry points in reciprocal space along the main path utilized for the data analysis shown in blue.

In the second setup, the crystal was realigned to the ($h$,0,$l$) scattering plane and placed in an Orange helium-flow cryostat and data were taken at temperatures of $T =$ 1.6, 60, 70, 72, 75, 80, and 100~K. Similarly to the first setup, data were acquired through sample rotation scans with a step size of 1$^\circ$ but with incoming energies of 5.50, and 5.63~meV and an angular setting of -75 and -71$^\circ$. All temperatures were measured using a monitor value of 100\,000 corresponding to 25 s/point at 5.5 meV, and twice this monitor value for 60~K.

\section{Results and discussion}
This section is split into three parts, firstly dealing with the magnetic field dependence of the directional and localized diffuse scattering intensity, secondly with its dependency on the scattering vector orthogonal to the frustrated triangular planes, {\em i.e.}\ the $l$ direction. The last section deals with 
%the magnon response to the external field and 
the additional intensity found along (0, 0, $l$) and how this advances the discussion of the magnetic ground state in \ymo.
\subsection{Magnetic field and energy dependence}
\begin{figure*}[htbp] % Figure generated by following python scripts combined in the power point file:
% C:\Users\lass_j\Documents\CAEMA2023\20222739_YMnO3\YMnO3_Old_Combined_ConstEMaps.py
% C:\Users\lass_j\Documents\CAEMA2023\20222739_YMnO3\100KConstantEnergyMap.py
% C:\Users\lass_j\Documents\CAEMA2023\20222739_YMnO3\Treasuremap2.py
% C:\Users\lass_j\Documents\CAEMA2023\20222739_YMnO3\figures\old\Fig2.pptx and saved as png
    %\includegraphics[width=0.75\linewidth]{Figures/Fig2_v4_part1.png}
    \includesvg[inkscapelatex=false]{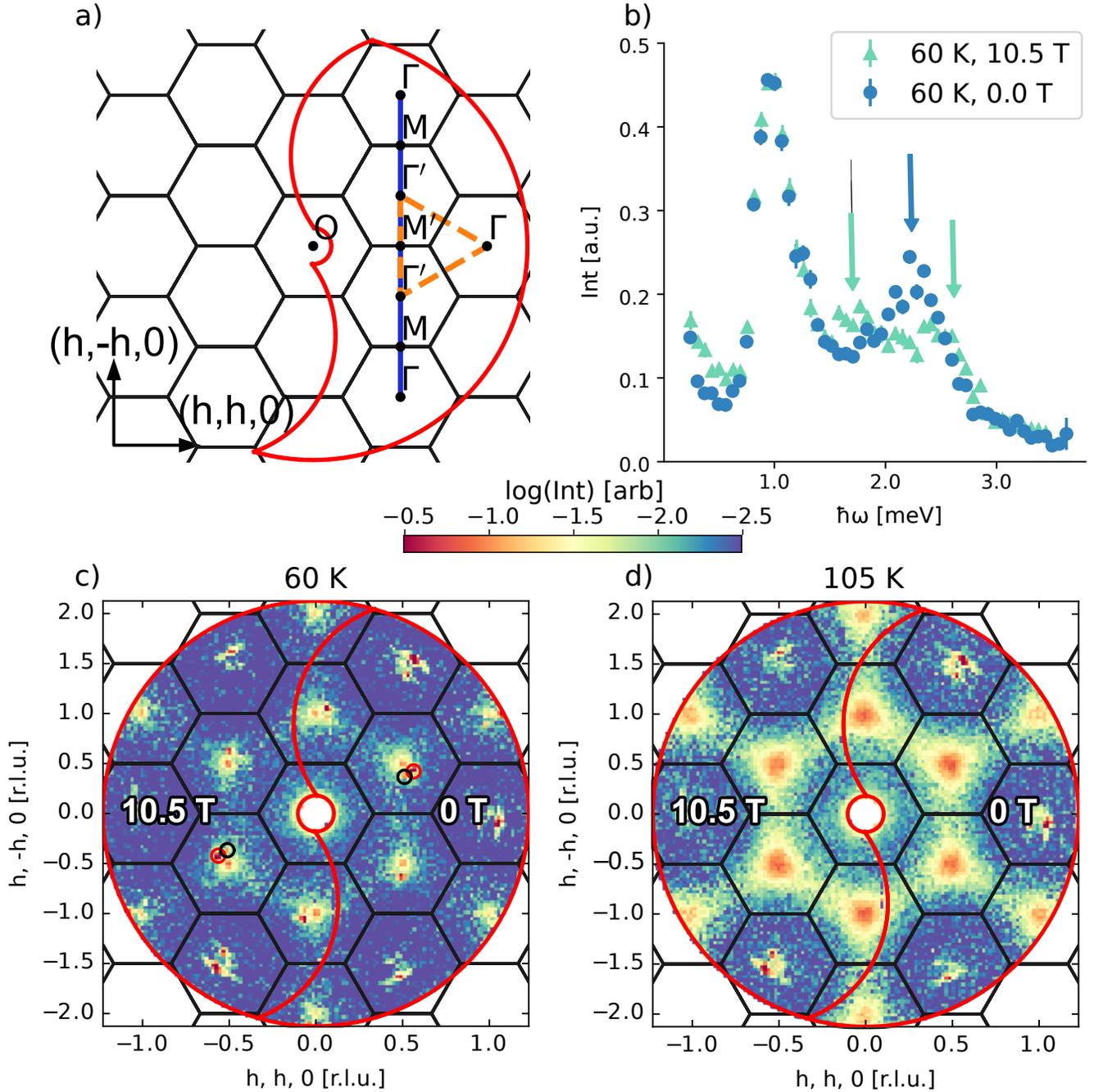}
    \caption{(a) Overview of reciprocal lattice coverage in the ($h$,$k$,0) plane of \ymo at $\hbar\omega = $ 1.0~meV. $\mathrm{O}$ is the origin and $\Gamma$ the position of allowed nuclear Bragg peaks, while $\Gamma'$ is the position of the magnetic Bragg peaks. $\mathrm{M}$ and $\mathrm{M}^\prime$ are on the two different zone boundaries. The boundary for data coverage at $\hbar\omega = 1.0$~meV is shown by the red curve while the orange dashed path correspond to the classical spin dynamics simulations.
    (b) Energy dependence on the external field for the spectrum at (1, 0, 0), within the ordered phase where the upper magnon splits into two (shown by the arrows).
    (c-d)
    Logarithmic intensity maps across the ($h$,$k$,0) plane for energy transfers of $\hbar \omega = 1.0\pm0.1$~meV for 60~K and 105~K, respectively. For both maps, B = 10.5~T and 0~T at the same temperature are shown within the same plot separated by the red outlines plotted to the left and right. This corresponds to a constant energy map below the spin wave in the ordered phase. The Currat-Axe spurions originating from ($\pm$1, 0, 0) from one set of $\Gamma^\prime$ points are encircled by red and black. 
    }
    \label{fig:Qmap_BZ_p1}
\end{figure*}

\begin{figure*}[htbp] % Figure generated by following python scripts combined in the power point file:
% C:\Users\lass_j\Documents\CAEMA2023\20222739_YMnO3\figures\old\Fig2.pptx and saved as png
    %\includegraphics[width=0.75\linewidth]{Figures/Fig2_v4_part2.png}
    \includesvg[inkscapelatex=false]{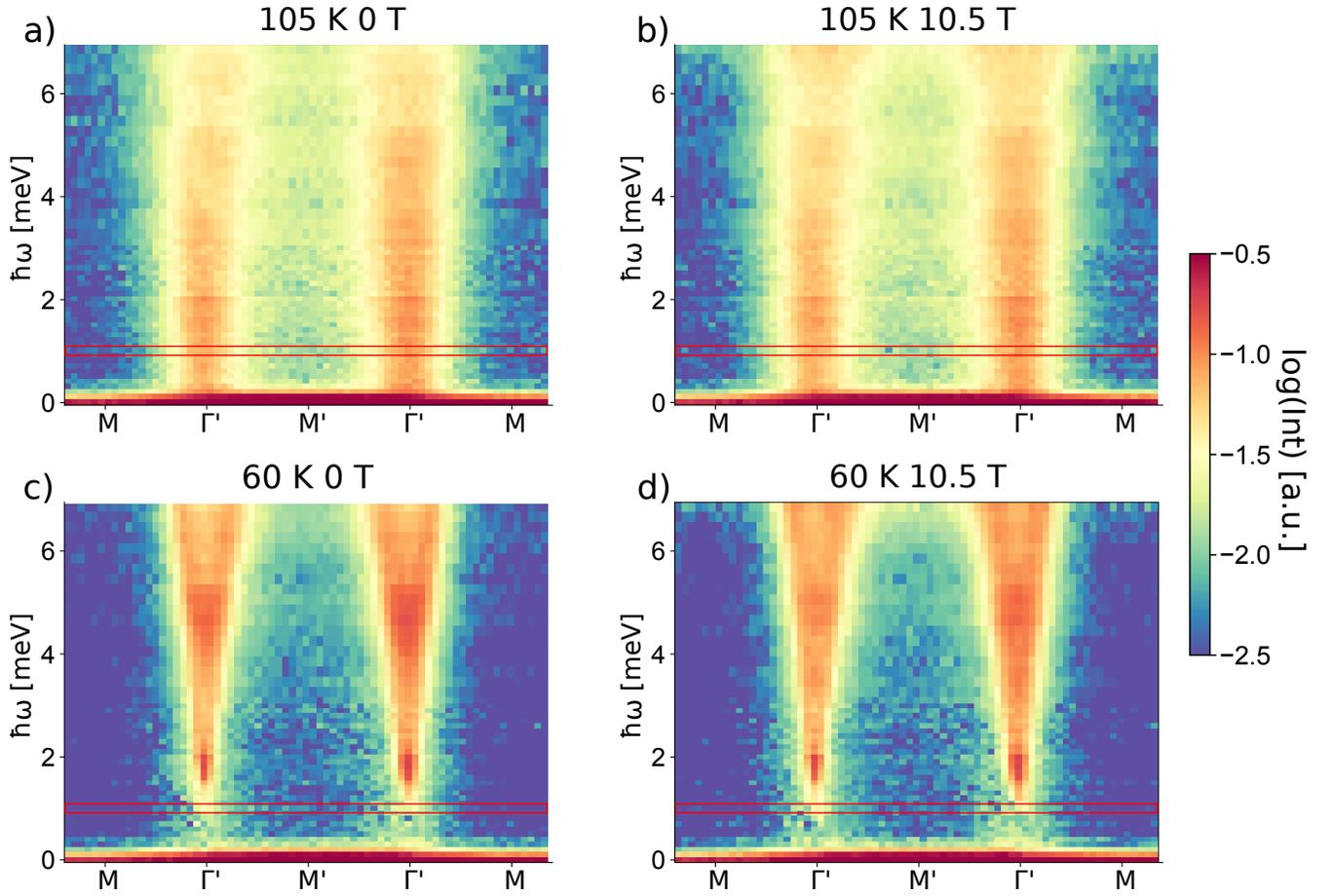}
    \caption{(a-b)
    Logarithmic intensity maps as function of energy transfer along the blue path shown in Fig.~\ref{fig:Qmap_BZ_p1}(a), with an orthogonal integration width of $\pm$0.15 r.l.u. for all both temperatures, 60 K and 105 K, in zero and high field. The horizontal line signifies the constant energy cuts corresponding to Figs.~\ref{fig:Qmap_BZ_p1}(c) and (d). Notice that the Currat-Axe spurions are masked out. 
    All intensities are plotted with the same color scale.
    }
    \label{fig:Qmap_BZ_p2}
\end{figure*}

%One aspect not touched upon in previous literature is the possibility of the diffuse magnetic excitations being dependent on external magnetic fields. Specifically, an applied field along the crystallographic $c$ direction could serve to induce a spin canting with a potential impact on the signals.
Data taken in zero field and at 10.5~T are shown in Figs.~\ref{fig:Qmap_BZ_p1} and \ref{fig:Qmap_BZ_p2}. The the coverage of the measured data at an energy transfer of $\hbar\omega = 1.0 \pm 0.1$~meV is shown in Fig.~\ref{fig:Qmap_BZ_p1}~(a) together with the nomenclature for high symmetry points in reciprocal space along the main path utilized for the data analysis highlighted in blue. Fig.~\ref{fig:Qmap_BZ_p1}~(b) shows the energy dependence of the scattering for the 60 K data in zero field and 10.5~T with arrows highlighting the spin wave positions.

Figs.~\ref{fig:Qmap_BZ_p1}~(c) and (d) show constant-energy cuts through $\hbar\omega = 1.0$~meV $\pm$ 0.1~meV for the temperatures of 105~K and 60~K, in zero field and 10.5 T. Data taken in the ordered phase were hampered by presence of spurious Bragg scattering appearing as inelastic signals, described by Currat and Axe \cite{Shirane2013}, also known as accidental scattering. In the plots, this spurious scattering is highlighted by red and black circles for the magnetic peaks at ($\pm$1, 0, 0). For lower energy transfers, the locations of these spurious features move close to the Bragg peak position, making a reliable determination of the scattering intensity at the $\Gamma^\prime$ point difficult. Temperatures of 60 and 105~K were chosen as they are well below and above $T_N = 72$~K. In the case of normal critical phenomena, expected due to the second order nature of the phase transition, critical scattering should hardly be visible at these temperatures.\cite{CollinsBook89} In the constant energy data in Fig.~\ref{fig:Qmap_BZ_p1}(c-d), we see a series of filled circle-like shapes with a slight triangular distortion, centered at the magnetic ordering vectors, $\Gamma^\prime$. Between these lines of directional, diffuse scattering are seen to form a hexagonal pattern, along the path $\Gamma^\prime$-$\mathrm{M}^\prime$-$\Gamma^\prime$. This scattering intensity is clearly seen to persist at 60~K, as also previously reported,\cite{Janas2021} but with a reduced intensity and width.

To quantify the effect of applied magnetic field on the diffuse continuum excitations, we perform a series of cuts along a single direction in reciprocal space (the blue line in Fig.~\ref{fig:Qmap_BZ_p1}~(a)) resulting in the data shown in Figs.~\ref{fig:Qmap_BZ_p2}~(a-d). They show the main cuts for the four different settings, {\em i.e.}\ temperatures of 105~K and 60~K, top and bottom, respectively, and with magnetic fields of zero (left) and 10.5~T (right). The effect on the spin wave observed in Fig.~\ref{fig:Qmap_BZ_p2}~(d) is consistent with the behaviour seen in Fig.~2 of Ref.~\onlinecite{Holm2018}, %with an energy broadening (splitting) of the spin wave gap at the $\Gamma^\prime$ for the 4~meV spin wave while the 1.5~meV spin wave remains unchanged. 
%As also found in Ref.~\onlinecite{Janas2021}, 
with the 60~K data in zero field showing clear spin waves at the $\Gamma^\prime$ points with energy gaps of $\sim$ 4.0~meV and 1.5~meV, below which a diffuse rod of scattering is present. Applying an external magnetic field along the $c$ axis changes the spin wave spectrum, where the higher spin wave branch splits in two, as already found in Ref.~\onlinecite{Holm2018}. However, the field does not appear to alter the diffuse excitations. At a temperature of 105~K, the spin waves are no longer present, and only the broad diffuse continuum excitations remain. These are present as rods of scattering at $\Gamma^\prime$, which are connected by a sheet of diffuse scattering extending through $M^\prime$. The diffuse excitations can be tracked up to at least our measurement range of 6.9~meV, broadening slightly as the energy transfer increases.% 

\begin{figure*}
% Generated by C:\Users\lass_j\Documents\CAEMA2023\20222739_YMnO3\OldDataArticleFigures.py 18/10-23 JL, V2: 09/11-23 JL
% Collected in C:\Users\lass_j\Documents\CAEMA2023\20222739_YMnO3\old\Fig3_v3.pptx
     \centering
         \includesvg[inkscapelatex=false]{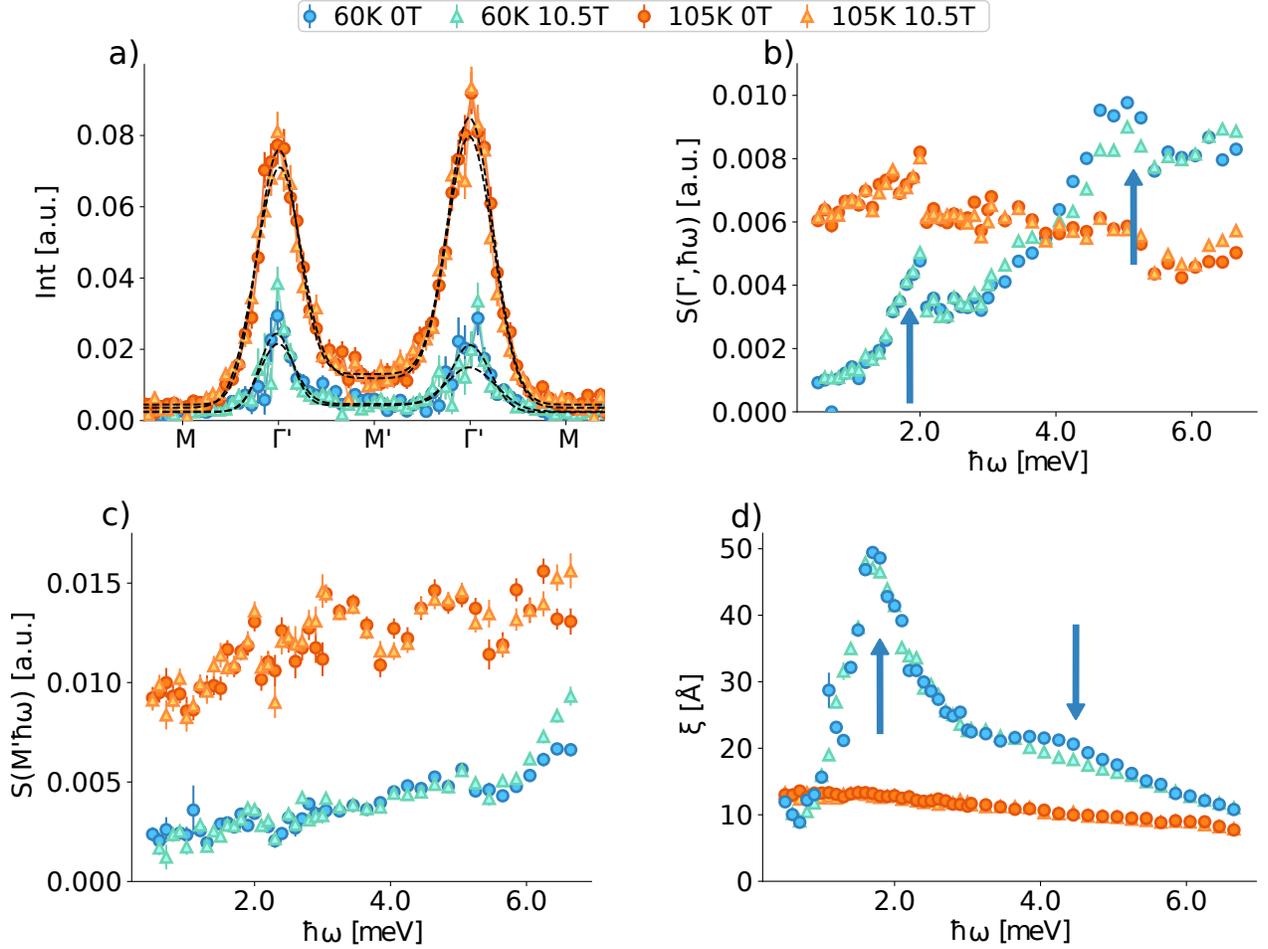}%\includegraphics[width=0.85\textwidth]{Figures/Fig3_v3.png}
    \caption{
    (a) Example of cuts along the signified blue line in Fig.~\ref{fig:Qmap_BZ_p1} at an energy transfer of 1.0 $\pm$0.1~meV,  at 60~K and 105~K, taken at zero field and at 10.5~T. The black, dashed line shows the corresponding fits.% Should this be improved? 
    (b-d)
    Fitted parameters from Eq.~\eqref{eq:fit} on constant-energy cuts of the plots in Figs.~\ref{fig:Qmap_BZ_p2}(a-d). (b) Integrated intensity at the $\Gamma^\prime$ points; (c) Peak intensity at the $\mathrm{M}^\prime$ point; 
    (d) Characteristic length, $\xi$, calculated by Eq.~\eqref{eq:xiDefinition}. Blue arrows in figures (b) and (d) signify the location of the magnon. }
        \label{fig:bridgefits}
\end{figure*}

Each 2D cut in Figs.~\ref{fig:Qmap_BZ_p2}~(a-d) is further sliced into constant 1D energy cuts with an integration width of 0.1~meV for energy transfers below 3.0~meV and 0.2~meV above. These cuts are fitted by a function consisting of a combination of two Gaussian functions for the signal at the $\Gamma^\prime$ points and a double-sided Sigmoid function for the signal at the $\mathrm{M}^\prime$ point, {\em i.e.}\ 
\begin{align}
I = &A_1~\exp{-\frac{(\mu_1-x)^2}{2\sigma_1^2}}+A_2~\exp{-\frac{(\mu_2-x)^2}{2\sigma_2^2}}+\notag \\
&2~S(\mathrm{M}^\prime,\hbar\omega)\left(\frac{1}{1+\exp{\frac{x-\mu_1}{\sigma_B}}}+\frac{1}{1+\exp{\frac{\mu_2-x}{\sigma_B}}}\right)+B.\label{eq:fit}%
\end{align}
Here, the Sigmoid transition width has been chosen to be linked to the Gaussian widths through $\sigma_B = (\sigma_1+\sigma_2)/8$, {\em i.e.}\ half the mean width. This choice for the fitting function is further discussed in Appendix~\ref{app:Fitting}. $S(\mathrm{M}^\prime,\hbar\omega)$ is the intensity at the $\mathrm{M}^\prime$ point at the energy transfer $\hbar\omega$, while $A_i$, $\mu_i$, and $\sigma_i$ are the intensities, centres, and widths of the $i^{\mathrm{th}}$ Gaussian and $B$ is a constant background, that is the mean of $A_i$ is the intensity at the $\Gamma^\prime$ point denoted $S(\Gamma^\prime,\hbar\omega)$. This combined function has been found to empirically describe the intensity across all temperatures and magnetic fields (as shown by the example in Fig.~\ref{fig:bridgefits}~(a)), except for energy transfers dominated by the magnon dispersion. In such cases, the two Gaussian functions mainly reflect the intensity of the magnon at $\Gamma^\prime$ instead of the desired diffuse scattering and the width is reduced due to the sharp nature of the long-lived magnon. However, the intensity at the $\mathrm{M}^\prime$ point between the magnon branches continues to be reliable.

The resulting fitting parameters as a function of energy transfer are shown in Figs.~\ref{fig:bridgefits}~(b) and (c) with an example of the fit at 1.0 meV in Fig.~\ref{fig:bridgefits}~(a). In addition, panel (d) shows the characteristic length, $\xi$, calculated from the fitted Gaussian width at $\Gamma^\prime$ and converted to the corresponding Lorentzian width, in order to adhere to the data in Ref.~\onlinecite{Janas2021}, as 
\begin{equation}
\xi = \frac{\sqrt{2\mathrm{ln}\left(2\right)}}{2\sigma} \label{eq:xiDefinition}.
\end{equation}
Fig.~\ref{fig:bridgefits}b shows that the intensity in the ordered phase (60~K) is consistent between the two fields up to the onset of the second magnon, $\hbar \omega \sim 4$~meV, where a field-induced change is found as expected. This effect is difficult to see from the color maps in Figs.~\ref{fig:Qmap_BZ_p2}. In the paramagnetic phase (105~K), the integrated intensity at $\Gamma^\prime$ decreases for increasing energy transfer but is unaffected by the magnetic field. The peak intensity at the $\mathrm{M}^\prime$ point, Fig.~\ref{fig:bridgefits}~(c), is wholly unaffected by the magnetic field, but the intensity slightly increases with increasing energy transfer, both in the ordered and paramagnetic phases. In the ordered phase, the magnon dominates the intensities at the $\Gamma^\prime$ point in the energy window from around 1.4 to about 6 meV, and for energy transfers below 0.7 meV the presence of the Currat-Axe spurions renders the intensity unreliable. Thus, the best estimate of the characteristic length at 60 K is to be made at 1 meV, consistent with previous results\cite{Janas2021}, where it takes on the value of $\sim$ 19~Å. This characteristic length is significantly smaller than the estimated instrument resolution of $2\pi/0.02$~Å$^{-1}~\approx320$~Å; the resolution is further discussed in Appendix~\ref{app:Resolution}.

The fitted parameters in Figs.~\ref{fig:bridgefits}~(b) and (c) show that the magnetic field only affects the mean integrated intensity above an energy transfer $\Delta E \sim 3$~meV, and only for the 60~K data at $\Gamma^{\prime}$ and the characteristic length through the excitation width. This is due to the presence of the magnon dispersion which is vastly more intense than the diffuse $\Gamma^\prime$ excitation. However, below the magnon gap the integrated intensity is unchanged, meaning that the diffuse continuum excitation is unaffected by magnetic field. The integrated intensity at the $\mathrm{\Gamma}^\prime$ point in the 105~K data does show a slight decrease for increasing energy transfers with the opposite effect seen for the $\mathrm{M}^\prime$ point. As for the intensity at both points, the 60~K data are more intense than the 105~K in agreement with previous data.\cite{Janas2021}

From the measurements in the ($h$, $k$, 0) scattering plane it is seen that both the diffuse excitations at the $\Gamma^\prime$ and $\mathrm{M}^\prime$ points extend from the elastic line all the way up to an energy transfer of at least 6.9~meV with a slight intensity decrease at the $\Gamma^\prime$ point and an intensity increase for the $\mathrm{M}^\prime$ point, {\em c.f.}\ Figs.~\ref{fig:bridgefits}(b) and (c).

At the $\mathrm{M}^\prime$ point the temperature dependence of both the low-energy intensity and the excitation width have previously been reported.\cite{Janas2021} These data show the intensity being lowest and width sharpest at 60~K, in agreement with our present results. Our new data, however, also show the energy dependence of this trend. The correlation length of the excitations at the $\Gamma^\prime$ point at 105~K decreases with increasing energy transfer, while the 60~K is heavily influenced by the magnon. Below the magnon, the intensity of the diffuse scattering continuously decreases until around 0.7 meV, where it takes an upturn due to the Currat-Axe spurions. Above the spin waves, {\em i.e.}\ above 6.3 meV, the 60~K data tend towards the 105~K data but with an increased width. Now, comparing the two temperatures, above the ordering temperature, $T_\mathrm{N}$, both parts of the diffuse signal are more intense and slightly sharper. Thus, the development of the 3D magnetic order does not directly impact the diffuse signal, but rather its intensity due to the sum rule. 

To summarize, our current data are consistent with previous results and extend the description of the diffuse scattering behaviour. All measures not dominated by the magnon are magnetic field independent. As a function of energy, the same overall trend of increasing intensity and reduction in correlation length is seen for both temperatures. This behaviour differs somewhat from the expected behaviour of a magnon spectrum, where intensities are expected to, except for the Bose-factor, remain constant, while the fitted correlation length only depend on resolution effects.

\subsection{Dimensionality of the excitation spectrum}
\begin{figure*}[tbhp]
    \includesvg[inkscapelatex=false]{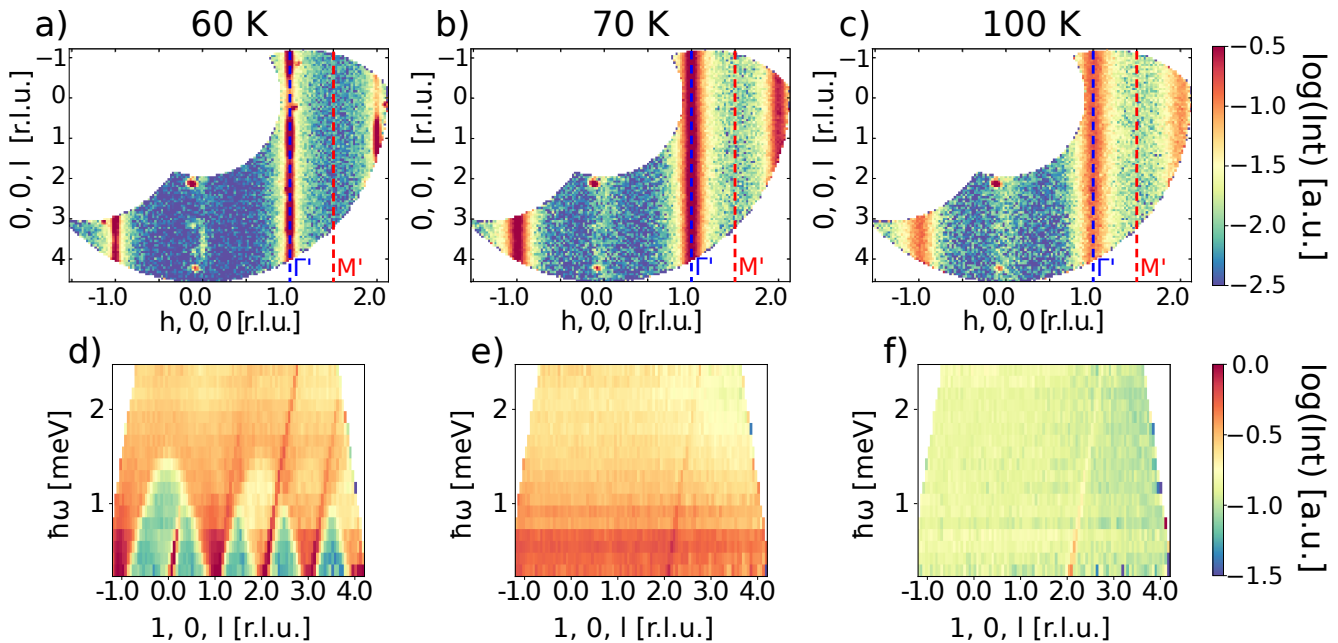}%\includegraphics[width=\linewidth]{Figures/fig4_v2.png}
    % from generated by: C:\Users\lass_j\Documents\CAEMA2023\20222739_YMnO3\100KConstantEnergyMap.py 19/09-23 using DiffuseMap60K_log_dotted_1meV_v2 and QEH1alongL60K_3
    
    \caption{
    (a-c)
    Logarithmic scattering intensity for constant energy maps in the ($h$,0,$l$) plane at $\hbar\omega = 1.0 \pm 0.1$~meV showing the diffuse scattering both at $\Gamma^\prime$ and $\mathrm{M}^\prime$. Dashed line signifies the diffuse scattering of the $\mathrm{M}^\prime$ red. %, with integrated 0.5 meV.
    %(top) 60~K, (center) 75~K, (bottom) 100~K.
    %Currat-Axe spurions from both magnetic and nuclear Bragg peaks have been masked out at all temperatures.
    (d-f)
    Logarithmic scattering intensity maps along ($h$,0,$l$) and $\hbar\omega$ at $h=1$, corresponding to the dashed $\Gamma^\prime$ line, with an integration width of 0.2 r.l.u. Diagonal signals  are the Currat-Axe spurions, which are masked out in the 1D cuts.
    }
    \label{fig:ConstantEnergyH0L}
\end{figure*}
Next we address the question of the nature of the diffuse magnetic scattering. It seems plausible that the signal arises from coupled fluctuations of the Mn$^{3+}$ moments within the triangular ($a$,$b$) plane. However, it remains unresolved if these fluctuations are coupled to the three-dimensional magnetic structure appearing below $T_{\rm N}$. 
%One main unanswered question from previous measurements remains: does the diffuse magnetic scattering originate solely from coupled fluctuations of the Mn moments within the triangular (a,b) plane, or is it also coupled to the three-dimensional structure so that it would track {\em e.g.}\ the phase transition to long-range order at $T_{\rm N}$? 
To resolve this, we investigated the dependence of the signal as function of $l$ within the ($h$,0,$l$) scattering plane. In addition, a possible dependence on the 3D ordering was investigated by measuring at various temperatures across $T_{\rm N}$.
%capturing the low energy transfer excitations. 
Three representative temperatures are shown in in Fig.~\ref{fig:ConstantEnergyH0L} %  \ref{fig:Qmaps_hl} 
for $T =$ 60, 75, and 100~K, For other temperatures see in section.~\ref{app:H0LConstEPlots}.

Figs. \ref{fig:ConstantEnergyH0L}~(a-c) show constant-energy cuts through the $(h,0,l)$ plane at an energy transfer of 1~meV at 60, 75, and 100~K while Figs. \ref{fig:ConstantEnergyH0L}~(d-f) show $(q,\hbar\omega)$ intensity maps at the same temperatures. The sharp peaks in (a-c) as well as the diagonal stripes in (d-f) are Currat-Axe spurions, which are masked out in the following data treatment. 

In the magnetically ordered phase, {\em i.e.}\ at 60~K, the magnon signal is clearly seen. Its periodicity is particularly  visible in Fig.~\ref{fig:ConstantEnergyH0L}(d) where it is ungapped and strong at (1,0,$l$) for non-zero values of $l$ and has an energy gap of 1.5~meV at (1,0,0), which is consistent with the data taken in the $(h, k, 0)$ plane. Due to the finite integration width orthogonal to the cut and the relatively steep dispersion of the spin wave perpendicular to $l$, the spin waves in Fig.~\ref{fig:ConstantEnergyH0L}~(d) do not appear to have clear upwards boundaries. At 75~K and 100~K (Fig.~\ref{fig:ConstantEnergyH0L}~(b-c) and (e-f)), no magnons are present and the intensity corresponds solely to the diffuse continuum excitations, which have no variation along the $l$ direction.  Rather, the signal is diffuse and gapless for all $l$-values. Very similar behaviours at all temperatures are also observed for $h = -1$ and $h = 2$ (data not shown). 

\begin{figure*}[hb]% Generated by C:\Users\lass_j\Documents\CAEMA2023\20222739_YMnO3\LComparison.py 26/09-23 V3: 09/11-23 JL
     \centering
     \includesvg[inkscapelatex=false]{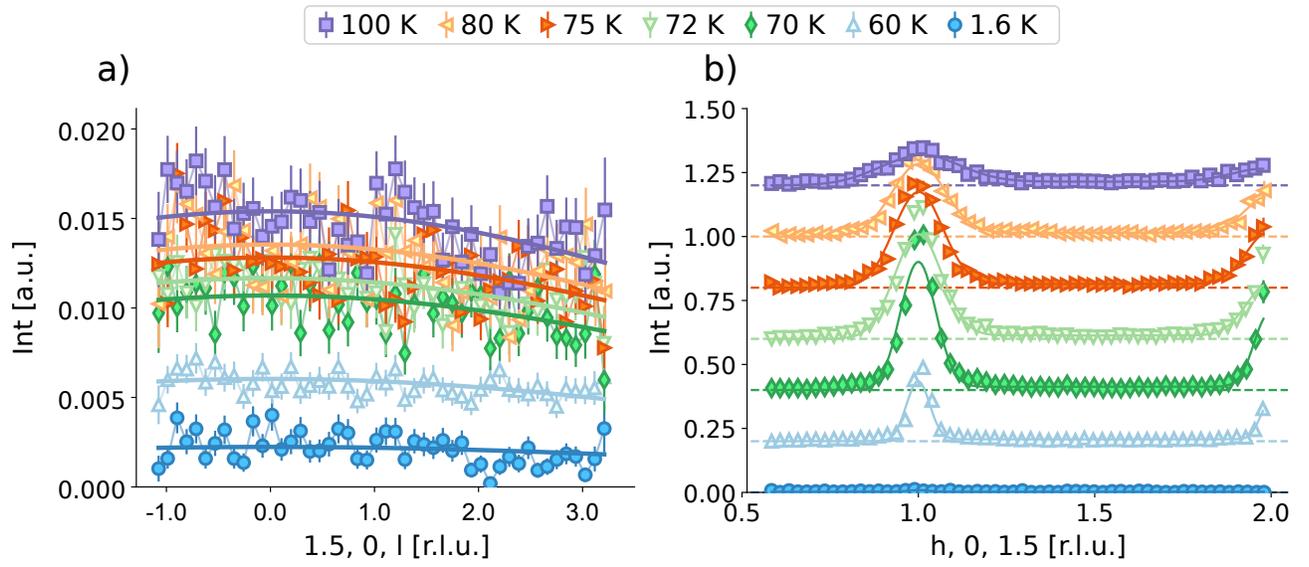}%\includegraphics[width=0.8\linewidth]{Figures/fig5_v3.png}
    \caption{The $l$-dependence of the intensity of inelastic data for (a) (1.5,0,$l$) and (b) ($h$,0,1.5) with an integration in energy from 0.9 to 1.1 meV and 0.2 r.l.u along $h$. The solid lines in (a) signify the magnetic form factor for Mn$^{3+}$, where their scale is optimized for each cut. The 6 data sets in (b) have been offset in intensity by each 0.2 units for clarity; the dashed line shows the individual zero points.}
    \label{fig:Qcuts_lAndh}
\end{figure*}

In order to investigate the $l$-dependence of the diffuse continuum excitations, we performed a cut for constant $h=1.5$, signified by the red dashed line in Figs.~\ref{fig:ConstantEnergyH0L}~(a-c), going through the $\mathrm{M}^\prime$ points at $\hbar\omega = 1.0\pm 0.1$~meV. This data is shown in Fig.~\ref{fig:Qcuts_lAndh}~(a). The chosen energy transfer is below the spin wave gap at $l = 0$ in the ordered phase for all measured temperatures. However, for all other integer values of $l$ the spin wave is visible and has thus been masked out. %This results in an absence of data in Fig.~\ref{fig:Qcuts_lAndh}~(a) around integer values of $l$. 
Comparing across all temperatures, the signal is close to being constant along $l$, with only a slight reduction towards higher $l$-values. This decrease is consistent with the magnetic form factor of Mn$^{3+}$ calculated using the tabulated values\cite{ILL} as explained in Appendix~\ref{App:magneticFormfactor}. This suggests that the structure factor of the diffuse continuum excitations for both values of $h$ - i.e. at the $\Gamma^\prime$ and $M^\prime$ points - are independent of $l$.

To further confirm the two-dimensionality and temperature independence, one-dimensional cuts were performed for half-integer values of $l$ along $h$ for an energy transfer of $1.0 \pm 0.1$ ~meV and integration width along $l$ of 0.09 r.l.u.\ to reduce the influence of the spin wave signal, see Fig.~\ref{fig:Qcuts_lAndh}~(b). Due to the varying $h$ coverage for the different values of $l$, the fitting function in Eq.~\eqref{eq:fit} is to be slightly adjusted such that the Sigmoid function always has its centre at the $\mathrm{M}^\prime$ points. Due to the limited coverage, the background parameter has been set to zero across all fits and the centre of the Gaussians are locked to the $\Gamma^\prime$ points.
% C:\Users\lass_j\Documents\CAEMA2023\20222739_YMnO3\ComparisonAcrossTemperatures_2.py

\begin{figure*}[htbp]  % Generated by C:\Users\lass_j\Documents\CAEMA2023\20222739_YMnO3\LComparison.py 26/09-23 V3: 09/11-23 JL
    \centering
    \includesvg[inkscapelatex=false]{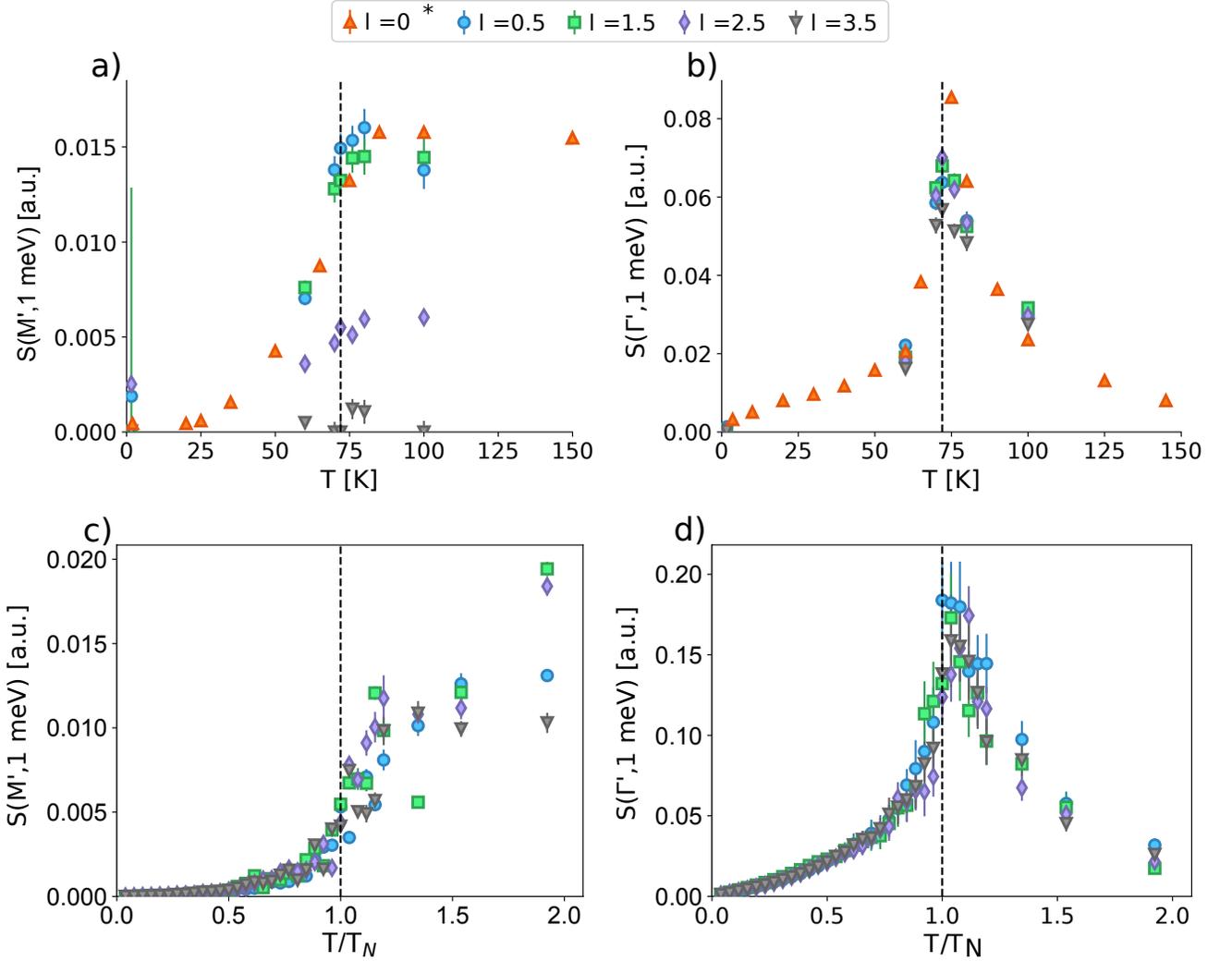}%\includegraphics[width=0.85\linewidth]{Figures/fig7.png}
    \caption{(a-b) Temperature dependence of the integrated intensity and peak intensity of the localized and directional diffuse scattering, respectively, from the cuts along ($h$,0,$l$) at half-integer values of $l$. \textbf{*}Data for $l = 0$ r.l.u. are taken from Ref.~\onlinecite{Janas2021} and rescaled to data at $l$ = 0.5, 1.5 r.l.u. and 80 K, utilizing the magnetic form factor. (c-d) Theoretical results for the integrated and peak intensities utilizing the spin dynamics simulations. The temperature dependence of the simulated data are plotted relative to the simulated $T_{\mathrm{N}}$ = 52~K. } %
   \label{fig:FittingParamsAlongL}
\end{figure*}

The scattering intensity at (1.5, 0, $l$), and at (1.0, 0, $l$) (not shown), follows the magnetic form factor of Mn$^{3+}$. % with only slight discrepancies due to the leaking of the magnon intensity into the signal are seen, Fig.~\ref{fig:Qcuts_lAndh}~(a).
In Fig.~\ref{fig:Qcuts_lAndh}~(b) the dependence of the $h$ component of the scattering vector is shown together with the above described fits. The integrated intensities and peak intensities corresponding to the $\Gamma^\prime$ and $\mathrm{M}^\prime$ points from these fits are shown in Fig.~\ref{fig:FittingParamsAlongL}. Specifically, in (a), the intensities decrease in accordance with the magnetic form factor as a function of $l$. The temperature trend is consistent with that reported in Ref.~\onlinecite{Janas2021}. To highlight this, the data for $l=0$ from  Ref.~\onlinecite{Janas2021} is plotted alongside the new data in Fig.~\ref{fig:FittingParamsAlongL}. It should be noted that for $l$ = 3.5 a discrepancy exists, where the $\mathrm{M}^\prime$ point intensity is significantly lower than expected. This is due to the limited coverage of the background found from the two $\Gamma^\prime$ points towards the $\mathrm{M}$ points at both ends of the fit. 
The integrated intensity at the $\Gamma^\prime$ points is unaffected by this background issue and follows the expected trend. Thus, it can be concluded that the magnetic spectrum at all temperatures only depends on $l$ through the variation in magnetic form factor. In particular, it rules out that the significant peak in intensity at the $\Gamma'$ point at $T_{\rm N}$, found in Ref.~\onlinecite{Janas2021}, originates from a dimensional cross-over from a low-temperature 3D ordering to a high-temperature 2D fluctuating regime. Thus, the signals at both $\Gamma'$ and $M'$ are of a fully two-dimensional nature.

We now compare the neutron data to simulated energy dispersions as calculated using spin dynamics simulations as implemented in the UppASD package.\cite{Antropov1995,Skubic2008,Etz2015,Eriksson2017} The largest practical size of 50x50x10 super-cell was simulated by cooling the randomly-initialized magnetic moments from the theoretically equivalence of 360~K down to 1~K as further described in Ref.~\onlinecite{Tosic2023}. Our chosen Hamiltonian model and parameters are from Ref.~\onlinecite{Tosic2022} and reproduce the magnetic ground-state $\Gamma_3$ of \ymo at temperatures well below T$_\mathrm{N}$ but produces a slightly shifted transition temperature of 52~K as compared to 72~K in the experiment. Thus, the temperature of the simulations is shown relative to the transition temperature. 

The dynamic response function was calculated along the path shown in orange in Fig.~\ref{fig:Qmap_BZ_p1}~(a), connecting two magnetic $\Gamma^\prime$ and one non-magnetic $\Gamma$ point through $\mathrm{M}^\prime$, $\mathrm{M}$, and $\mathrm{M}^\prime$ for $l$ = $\{0.5,1.5,2.5,3.5\}$. To find the expected scattering intensity from these response functions, the magnetic signal from the moments orthogonal to $\vec{q}$ is first extracted and the magnetic form factor for Mn$^{3+}$ is applied. This produces the relative intensities as would be measured by a neutron experiment. To concur with the experimental data, scattering intensities between 0.9 and 1.1~meV are integrated. Due to the simulation path, the directional diffuse scattering at $\mathrm{M}^\prime$ is found as the average intensity around the $\mathrm{M}^\prime$ point. The localized signal at $\Gamma^\prime$ is then found by performing a numerical integration across the $\Gamma^\prime$ point with the directional diffuse signal subtracted. 

The found intensities are plotted in Fig.~\ref{fig:FittingParamsAlongL}~(c-d) and show a quantitative agreement between the experimental and simulated data. The exact onset of the intensity at the $\mathrm{M}^\prime$ point going from low to high temperature differs between experiments and simulation, where the simulated intensities diverge faster. Above the phase transition, the intensity flattens within the plotted temperature range but despite the fact that the magnetic form factor has been taken into account when presenting the simulated data its dependence on $l$ seen in the experiment is not completely recovered. The same discrepancy is seen when comparing the intensity at the $\Gamma^\prime$ point where the divergence of the simulation is more abrupt compared to the experiment. These differences as well as the phase transition temperature is speculated to be due to a larger out of plane coupling or the introduction of impurities and stacking faults in the real world experiment compared to the idealized simulation. Further details on this simulation can be found in Ref.~\onlinecite{Tosic2023}.

Thus, due to the dependency of the experimental data on the $l$ component of the scattering vector only through the magnetic form factor and the replication of the general features through spin dynamic simulations it can be concluded that the diffuse signals originate from 
frustration in the $(a,b)$-plane and that they are in agreement with classical spin dynamics simulations using an approximate Hamiltonian of the system. 
%the frustrated in-plane magnetic Mn$^{3+}$ moments.

\subsection{Magnetic ground state from the magnon dispersions}\label{sec:spinwave}
\begin{figure*}[htb]
    \centering % Generated by C:\Users\lass_j\Documents\CAEMA2023\20222739_YMnO3\00LPlot.py 26/09-23 V3: 09/11-23 JL
    \includesvg[inkscapelatex=false]{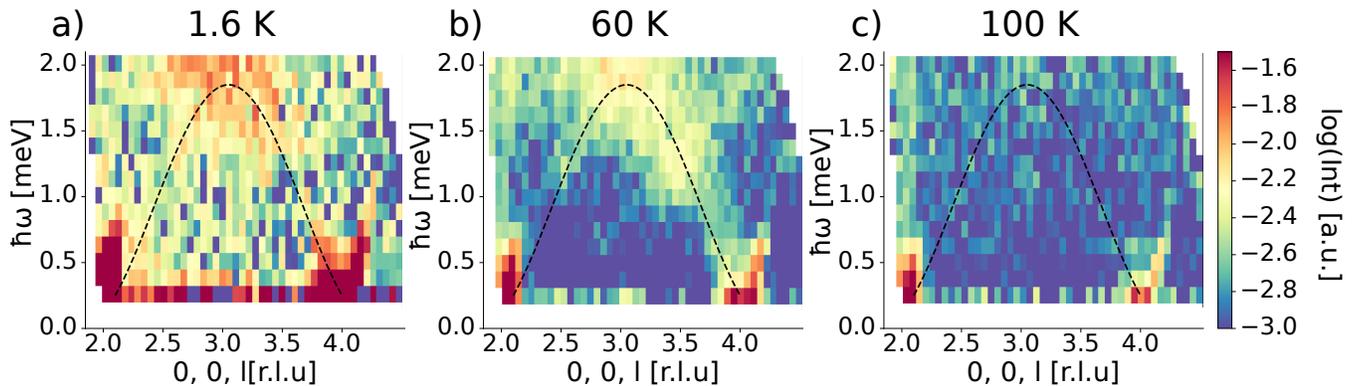}%\includegraphics[width=\linewidth]{Figures/fig6.png}
    \caption{Logarithmic intensity for $(\mathbf{q},\hbar\omega)$ maps along (0,0,$l$), integrated $\pm 0.04$~r.l.u along $h$ at (a) 1.6~K, (b), 60~K, and (c) 100~K with a guide to the eye capturing the signal at 60~K plotted by the black dashed line. Intensities have in all plots been rescaled by the inverse of the Bose-factor at their respective temperature.}\label{fig:QEMaps_0LPhonon}
\end{figure*}

The magnon dispersion in \ymo has already been measured extensively in the ($h$, $k$, 0) plane \cite{Holm2018, Oh2016, Chatterji2007} and to a smaller extent for non-zero $l$ values.\cite{Petit2008} The larger data coverage of our experiments extends the measurement range along the $l$ direction, and are consistent with the previous results in the ($h$, $k$, 0) plane. In those experiments the magnon spectrum was found to consist of two gapped low-lying spin waves with minima above the $\Gamma^\prime$ points and to extent to large energy transfers, between 10 and 15 meV. In addition, the upper of the two modes splits into two upon application of external magnetic field.

From the perspective of the irreducible magnetic representations, the Mn$^{3+}$ ions located on the 6c Wyckoff position of the P6$_3$cm (185) space group combined with the AFM ordering vector, four representations are allowed, $\Gamma_1$-$\Gamma_4$ shown in Fig.~\ref{fig:GroundStates}. From magnon calculations it was concluded that the proposed two possible ground state structures,\cite{Holm2018} $\Gamma_1$ and $\Gamma_3$, are consistent with all of our measurements in the $(h, k, 0)$ plane.

Inspecting the spectrum along (0, 0, $l$), plotted for 1.5, 60, and 100~K in Fig.~\ref{fig:QEMaps_0LPhonon}, a gapless mode for even, non-zero values of $l$ is present, which is most visible in (b), {\em i.e.}\ at 60~K. The positions of higher intensity correspond to allowed nuclear Bragg reflections and it may therefore be argued that the signal could originate from a phonon. However, the signal is almost absent in the 100~K data, where a phonon would have an increased intensity as compared to 60~K, as well as for low energy transfers at base temperature, {\em i.e.}\ 1.5~K. To highlight the intensity, the plots have been rescaled by the energy-dependent down-scattering Bose factor, $1-\exp{\hbar\omega/(k_BT)}$. 

Comparing to the previously measured magnon intensity at ($h$, 0, 0) (see Figs.~\ref{fig:Qmap_BZ_p2} (c) and (d)), the signal along (0,0,$l$) is two to three orders of magnitude weaker. From spin wave calculations, see App.~\ref{App:SpinW}, a magnetic signal is expected to exist along (0, 0, $l$) with a significantly lower scattering intensity as compared to the spin waves along ($h$, 0, 0), around four orders of magnitude weaker. Interestingly, the two proposed solutions, $\Gamma_1$ and $\Gamma_3$, predict different periodicity of the signal along (0, 0, $l$) with $\Gamma_1$ having higher intensities for odd $l$ and $\Gamma_3$ for even $l$. This suggests that $\Gamma_1$ is consistent with the data. The main difference between the two proposed representations is the ferro- vs. antiferromagnetic ordering of the relative orientation between the nearest neighbour spins along the crystallographic $c$ direction for $\Gamma_1$ and $\Gamma_3$, respectively. Further, we note that the inter-planar couplings, {\em i.e.}\ couplings with a component along $c$, are 4-5 orders of magnitude weaker than the dominant in-plane nearest-neighbour AFM exchange. This is due to the spins interacting through both the super-exchange over the oxygen ions as well as the direct coupling between the empty Mn 3d states. In this respect, \ymo is very similar to LuMnO$_3$\cite{Das2014} as the electronic orbital occupancy of Lu$^{3+}$ is equivalent to that of Y$^{3+}$. The super-exchange favours an antiferromagnetic ordering while the direct coupling induces a ferromagnetic coupling. In the present case, the ferromagnetic coupling is much stronger and $\Gamma_3$ should be favoured. 

This does, however, seem to be in disagreement with the magnon spectrum measured along (0, 0, $l$) where $\Gamma_1$ best reproduces the data, albeit this scattering is several orders of magnitude weaker than along the orthogonal directions. The large difference in energy scale between FM in-plane coupling and the AFM out-of-plane coupling generates a quasi-flat magnetic energy surface where small changes in chemical composition or defect-related changes to the crystal structure can lead to a realization of the different magnetic ground states. This could explain the discrepancy between theory and the best fit of the inelastic neutron scattering data, as one can expect antiferromagnetically ordered moments to form between stacked domains of oppositely ordered $\Gamma_3$ states.

%\FloatBarrier
\section{Conclusion}
In summary, our experiments confirm and extend the findings of the two-dimensional diffuse fluctuations, which are present in an extended temperature range around the long range magnetic ordering temperature. Our results also extend the discussion of the magnetic ground state in \ymo.

The scattering continua are unaffected by an applied external magnetic field and show no dependence on scattering vector along the crystallographic $l$ direction, apart from the magnetic form factor. We confirm the two-dimensional nature of the fluctuations as previously suggested in Ref.~\onlinecite{Janas2021} and distinguish them from paramagnetic scattering and any possible leaking of intensity from the out-of-plane magnon due to the finite resolution orthogonal to the ($h$,$k$,0) plane. Furthermore, the scattering intensity localized on the magnetic Bragg peak positions, $\Gamma^\prime$, has a decreasing intensity as a function of energy transfer, while the directional diffuse scattering, centered around the Brillouin zone boundary point $\mathrm{M}^\prime$, remain constant. This confirms that the two signals, although likely connected, are of different nature. Both signals decrease similarly in intensity for increasing momentum transfer, consistent with the scattering following the magnetic form factor being two-dimensional and originating from the Mn$^{3+}$ ions. 

The temperature dependence of the continuous scattering intensity has been reproduced by atomistic spin dynamics simulations, which confirms its independence of the $l$ component of the momentum transfer. We thus confirm that the excitations originate from the strongly frustrated magnetic moments located in the triangular pattern without connection to neighbouring triangles along the crystallographic $c$ axis. This is in contrast to the magnon dispersion from the 3D magnetic ordering.

Through the spin wave spectra measured in the two scattering planes, the magnetic ground state is found to be consistent with $\Gamma_3$ in combination with either a spin canting or stacking faults resulting in the weak scattering observed along (0, 0, $l$). This is in agreement with the conclusions drawn from second generation harmonic experiments. \cite{Fiebig2000}

The observed diffuse fluctuation spectrum is distinctly different from the established theory of critical scattering originating from magnetic phase transitions both in dimensionality and temperature dependence. We believe that these detailed results will contribute to establishing a more detailed understanding of the peculiar directional diffuse, inelastic scattering observed in \ymo. 

%%%%%%%%%%%%%%
%\section{Outlook}

\ymo is one example of a frustrated magnetic system, where two magnetic signals coexists with completely different temperature dependence, scattering vector, and magnetic field behaviour. The origin of this two component behaviour is believed to be the 2D triangular frustration of the localized magnetic moments in the $(a,b)$-plane which suppresses the long range order from the Curie-Weiss temperature of $\sim$ 500~K to $T_N$ = 72~K. It further leads to the fully in-plane nature of the directional diffuse scattering. The separate temperature and scattering vector dependence of the coexisting magnetic signals might be taken to be a more general feature of classical spin liquids suggesting that other systems within this category would be magnetically frustrated materials with large spin clusters, that order well below the Curie-Weiss temperature.

\begin{acknowledgments}
This work was supported by the Danish Agency for Science, Technology, and Innovation through the instrument center DanScatt; Grant number 7129-00006B.
This work was funded by the European Research Council (ERC) under the European Union’s Horizon 2020 research and innovation program project HERO grant agreement No. 810451.
The neutron experiments were performed at the SINQ neutron source, Paul Scherrer Institute, Switzerland. We would like to thank Peter S. Beck and Thomas B. Hansen for their help in performing the first experiment. Computational resources were provided by ETH Zürich and the Swiss National Supercomputing center, project ID s889. We thank A. Kreisel for providing the calculation of the magnetic dispersions for the four ground states used in Ref.~\onlinecite{Holm2018}. Further, we thank Prof. Dr. N. Spaldin for rewarding discussions and Prof. Dr. C. Niedermayer for his help with the manuscript.
\end{acknowledgments}
\bibliography{bib}

\newpage
\onecolumngrid
\appendix
\renewcommand{\figurename}{App Fig.}
\FloatBarrier\newpage

%\section{Things to be pushed in}
%Sample mass, and why it got decreased

\section{Fitting details}\label{app:Fitting}
The fitting function defined in eq.~\eqref{eq:fit} is an empirical function chosen based on the signals for the two cuts in the ($h$,$k$,0) plane drawn by the blue line in Fig.~\ref{fig:Qmap_BZ_p1}(a) and along constant $l$ lines in the ($h$,0,$l$) plane. For both cuts, the localized continuum scattering can be captured for all $\Gamma^\prime$ points located within the cuts. This number varies depending on the $h$ value. In addition, the directional diffuse scattering through the $\mathrm{M}^\prime$ is captured as well. 
For the cuts in the ($h$,$k$,0) plane all parameters are allowed to vary, although the background is constrained to $B>0$. For the ($h$,0,$l$) plane, the background is constrained to $B=0$ and the location of the Gaussians modeling the $\Gamma^\prime$ excitations are locked to their integer positions. These restrictions are required for the fit to converge to reasonable solutions throughout the scattering plane and results in the directional diffuse scattering being determined the clearest. Furthermore, depending on the value of $l$, the Sigmoid function is to be changed carefully such that it connects two $\Gamma^\prime$ points but not a $\Gamma^\prime$ point with a non-magnetic $\Gamma$ point. Only in the case of a constant $l$ cut with $h$ = 2~r.l.u can a single Gaussian be fitted due to the coverage of the data, see Fig.~\ref{fig:ConstantEnergyH0L} at e.g. 60~K. 

\begin{figure*}[htb!] 
    \begin{subfigure}[b]{0.45\textwidth}
         \centering
         \includegraphics[width=\textwidth]{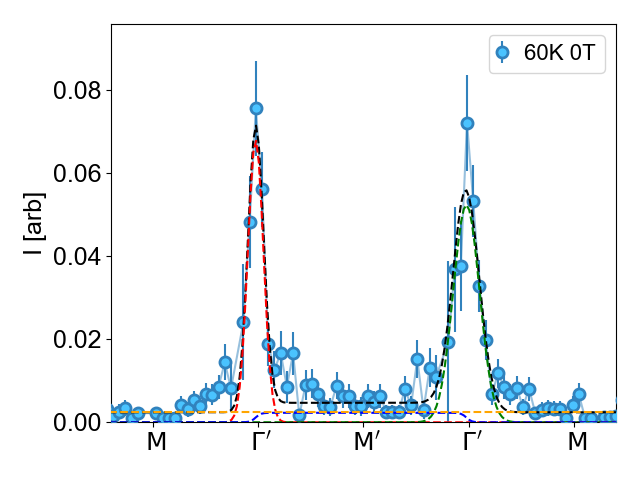}
         \caption{}%$\Delta E$ = 0.4 meV}
     \end{subfigure}%
     \begin{subfigure}[b]{0.45\textwidth}
         \centering
         \includegraphics[width=\textwidth]{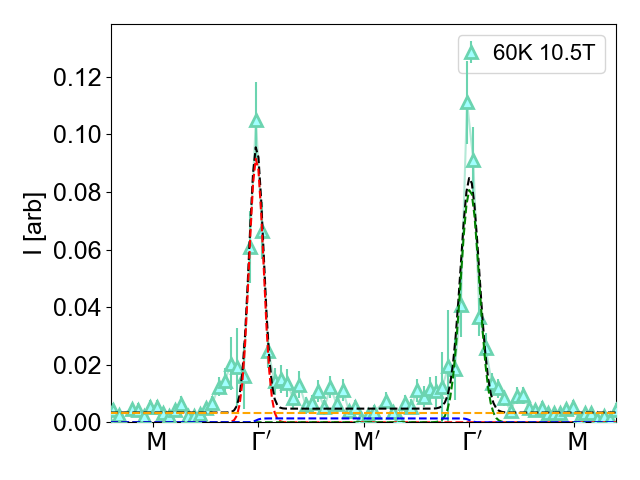}
         \caption{}%$\Delta E$ = 0.4 meV}
     \end{subfigure}
     \begin{subfigure}[b]{0.45\textwidth}
         \centering
         \includegraphics[width=\textwidth]{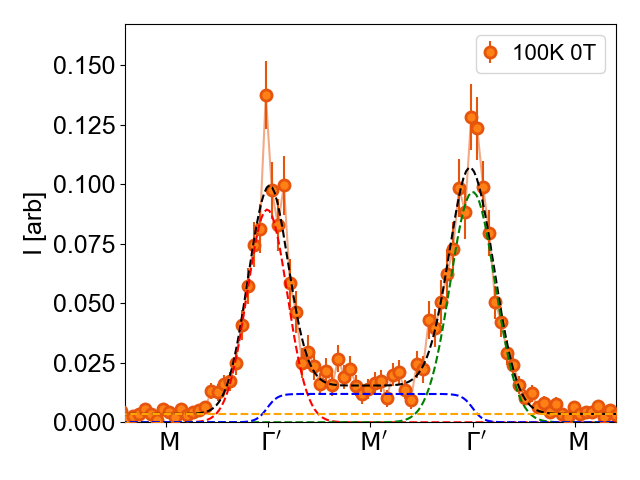}
         \caption{}%$\Delta E$ = 0.4 meV}
     \end{subfigure}%
     \begin{subfigure}[b]{0.45\textwidth}
         \centering
         \includegraphics[width=\textwidth]{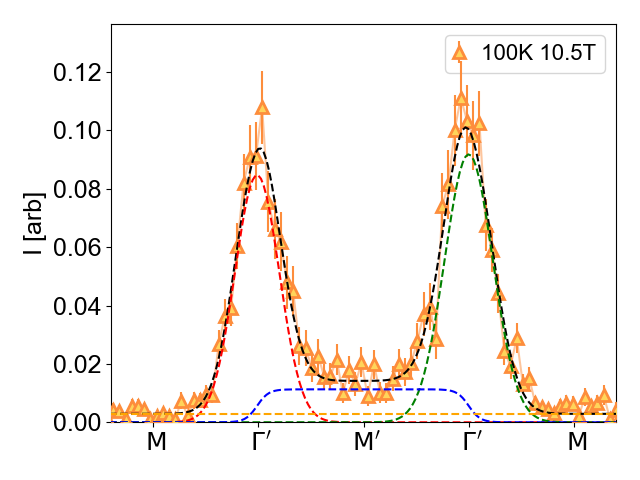}
         \caption{}%$\Delta E$ = 0.4 meV}
     \end{subfigure}
        \caption{Detailed view of fitting function along the cut specified in Fig.~\ref{fig:Qmap_BZ_p1}(a) for all four settings showing the resulting fit(black) of the data at $\Delta E$ = 0.7~meV~$\pm$~0.05~meV together with the individual contributions as dashed lines from the two Gaussians({\color{AppRed}red} and {\color{AppGreen}green}), the Sigmoid({\color{AppBlue}blue}) and the constant background({\color{AppOrange}orange}).}
        \label{fig:FitExample0p4}
\end{figure*}

\section{Instrument resolution}\label{app:Resolution}
In the analysis of the located and directional scattering in \ymo, the widths of these excitations are of high importance. In order to faithfully determine their inherent values, the broadening due to the experimental setup has to be well known. Neutron scattering, and especially inelastic, instruments are built balancing the flux and resolution, often times compromising one for an increase in the other. In the present case of CAMEA, the flexibility of the instrument is rather limited but the resolution has to some extent been characterized through theoretical calculations, i.e. through a modified Eckold-Sobolev calculation\cite{Eckold2014} which has been compared to well-established signals, in our case MnF$_2$ with the result of the calculations being slightly larger than what is observed in reality. Thus, the resolution calculations serve as an upper bound of the instrument broadening. 

\begin{figure*}[htb!]  % C:\Users\lass_j\Documents\CAEMA2023\20222739_YMnO3\YMnO3_Old_2021.py 27-09-23
    \centering
    \includegraphics[width=0.75\linewidth]{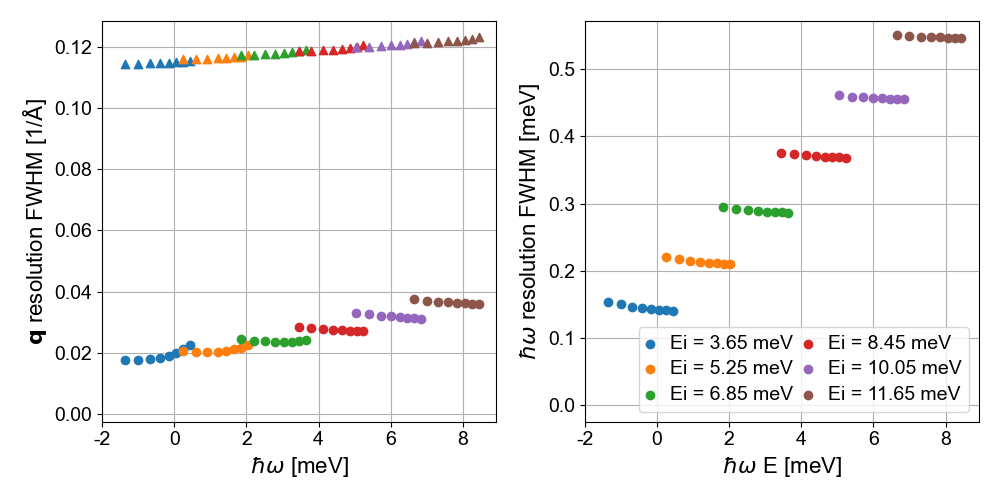}
    \caption{Calculated resolution for instrumental setup used for CAMEA experiment with the sample oriented in the ($h$,0,$l$) plane for all of the incoming energies. The $\mathbf{q}$ resolution along and orthogonal to the scattering vector is denoted by triangles and circles, respectively.}
    \label{fig:resolution}
\end{figure*}

The change of width of the diffuse continuum excitation at $\Gamma^\prime$ as a function of energy, plotted App.Fig.~\ref{fig:FitExample0p4}, increases slightly for increasing energies. This poses the natural question of whether it is intrinsic to the signal or due to the convolution with the instrumental resolution. The calculated resolutions for the setup used to measure \ymo utilizing the incoming energies of the experiment with the sample oriented in the ($h$,0,$l$) plane are plotted in App.Fig.~\ref{fig:resolution} at the $\mathrm{M}^\prime$ point. As the direction of the width defining the Gaussian is orthogonal to the scattering vector, the best estimate for the resolution is a FWHM of around 0.02 \AA$^{-1}$ which is converted to  0.017 \AA$^{-1}$  and comparing to the found widths if the diffuse scattering of maximally 20 \AA, or correspondingly 0.030 \AA$^{-1}$ it is seen that the resolution is around half the measured width. As the combined width of a folding of two Gaussian functions is $\sigma_\mathrm{tot} = \sqrt{\sigma_1^2+\sigma_2^2}$, the intrinsic width can be estimated to be 0.025 \AA$^{-1}$ or correspondingly 23.5 \AA, {\em i.e.} only slightly changed. As the instrument resolution does not change much as function of energy, the found behaviour is believed to be solely originating from the sample.

\FloatBarrier
\section{The \texorpdfstring{$l$}{l}-dependence of the scattering for all temperatures}\label{app:H0LConstEPlots}
\begin{figure*}[htb!] % C:\Users\lass_j\Documents\CAEMA2023\20222739_YMnO3\100KConstantEnergyMap.py 27-09-23
    \begin{subfigure}[b]{0.45\textwidth}
         \centering
         \includegraphics[width=\textwidth]{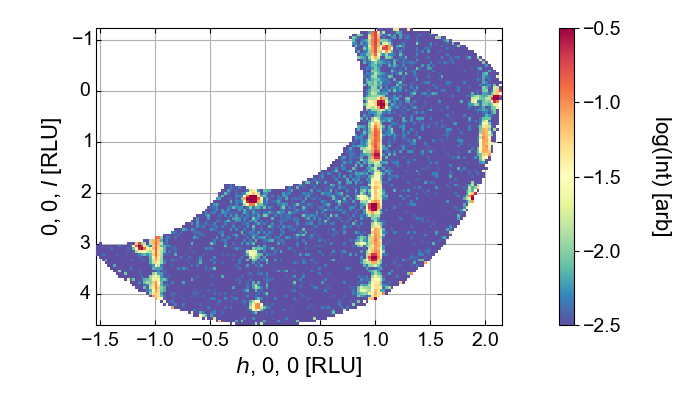}
         \caption{T = 1.6~K}%$\Delta E$ = 0.4 meV}
     \end{subfigure}%
     \begin{subfigure}[b]{0.45\textwidth}
         \centering
         \includegraphics[width=\textwidth]{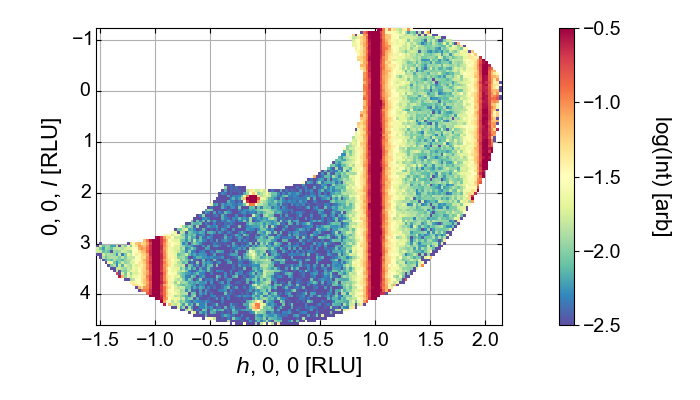}
         \caption{T = 70~K}%$\Delta E$ = 0.4 meV}
     \end{subfigure}
     \begin{subfigure}[b]{0.45\textwidth}
         \centering
         \includegraphics[width=\textwidth]{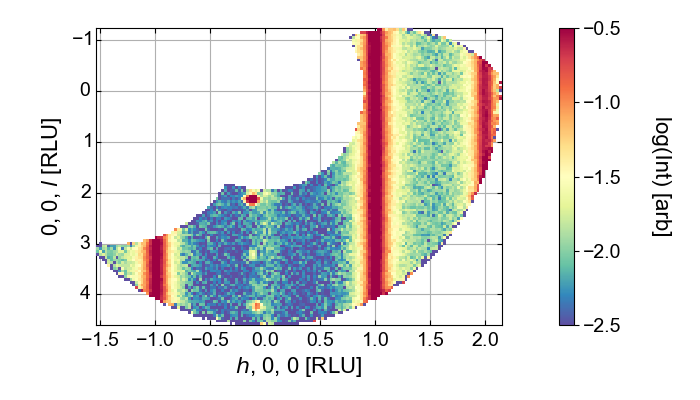}
         \caption{T = 72~K}%$\Delta E$ = 0.4 meV}
     \end{subfigure}%
     \begin{subfigure}[b]{0.45\textwidth}
         \centering
         \includegraphics[width=\textwidth]{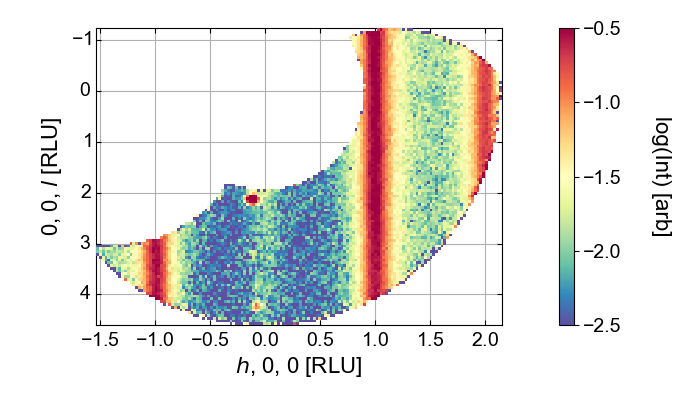}
         \caption{T = 80~K}%$\Delta E$ = 0.4 meV}
     \end{subfigure}
        \caption{Scattering intensity in the ($h$,0,$l$) plane at an energy transfer of 1~meV~$\pm$~0.25 meV extending the temperature range from Fig.~\ref{fig:ConstantEnergyH0L}.}
        \label{fig:app:Qmaps_hl}
\end{figure*}

For the setup with ($h$,0,$l$) in the plane, in addition to the three temperatures plotted in App.Fig.~\ref{fig:ConstantEnergyH0L}, four more temperatures have been measured and are plotted in Fig.~\ref{fig:app:Qmaps_hl}. %In contrast to the main figure, the Currat-Axe spurions are here plotted for clarity and it can be seen that, e.g.,  the spurions corresponding to the magnetic (0,0,2) Bragg point is faintly visible for T = 80~K due to critical fluctuations and increases in strength for lower temperatures. 
As a function of energy transfer, no dependency is found at any temperature above the phase transition. 

\FloatBarrier

\section{The magnetic form factor }\label{App:magneticFormfactor}
\begin{figure}[htb!] 
    \centering
    \includegraphics[width=0.45\textwidth]{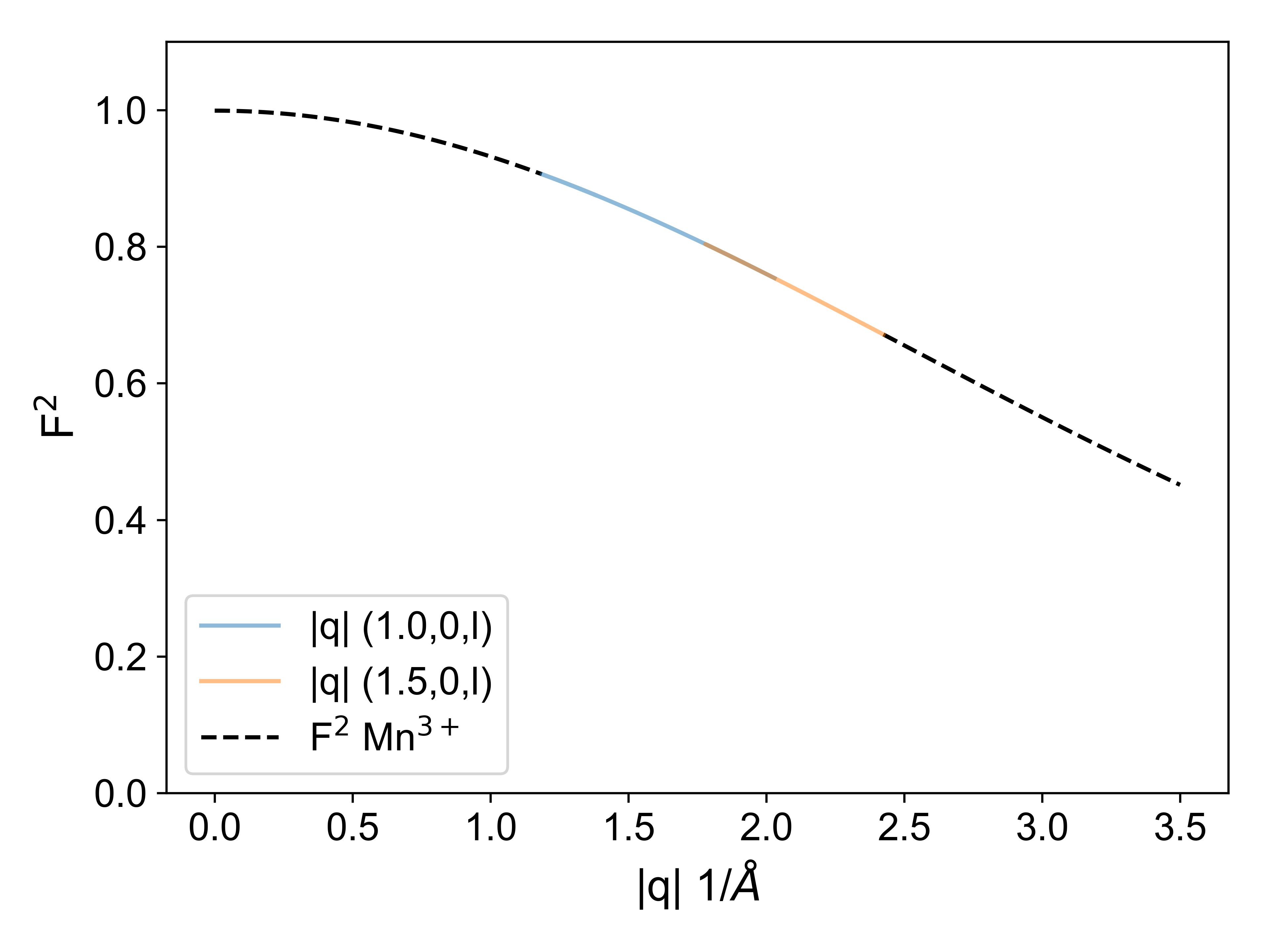}
    \caption{Square of the calculated magnetic form factor for Mn$^{3+}$ with the blue and orange curve signifying the $|q|$ coverage for a cut along (1.0,0,$l$) and (1.5,0,$l$).}
    \label{fig:app:MagneticFormFactor}
\end{figure}

The magnetic form factor of most magnetic ions are tabulated online\cite{ILL} and the expected decrease in scattering intensity from magnetic signals can be calculated for a given length of the scattering vector. For Mn$^{3+}$ in \ymo, the resulting expected intensity is plotted in App.Fig.~\ref{fig:app:MagneticFormFactor}, where the solid blue/orange line shows the length of the scattering vector for a cut along (1, 0, $l$)/(1.5, 0, $l$), with a minimal length of 1.78 1/Å and maximal length of 2.42 1/Å while the black dotted line traces out the form factor for an extended range. 

\section{Comparison of Magnon to previously published Spin Hamiltonian}\label{App:SpinW}
\begin{figure*}[htb!] %  C:\Users\lass_j\Documents\CAEMA2023\20222739_YMnO3\plotSPINWData.py after data generation using
                      % C:\Users\lass_j\Documents\YMnO3\YMnO3_sw.tar\sp_ph_ymno3_minimal\simple_JL.m   18/10-23
    \centering
    \begin{subfigure}[b]{0.32\textwidth}
    \includegraphics[width=\linewidth]{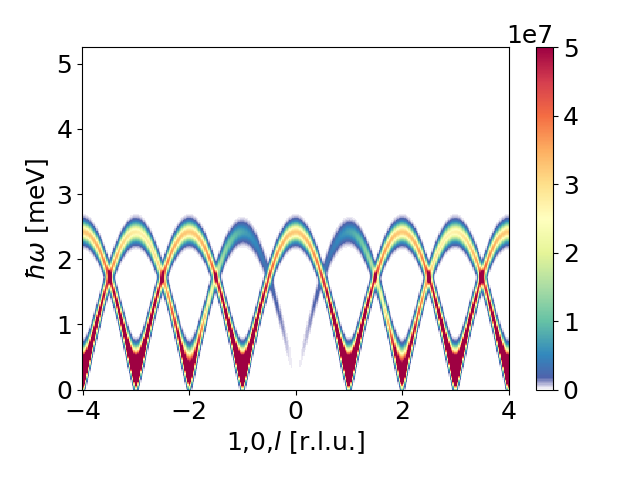}
    \caption{}
    \label{fig:app:spinw10lGamma1}
     \end{subfigure}%
     \begin{subfigure}[b]{0.32\textwidth}
    \includegraphics[width=\linewidth]{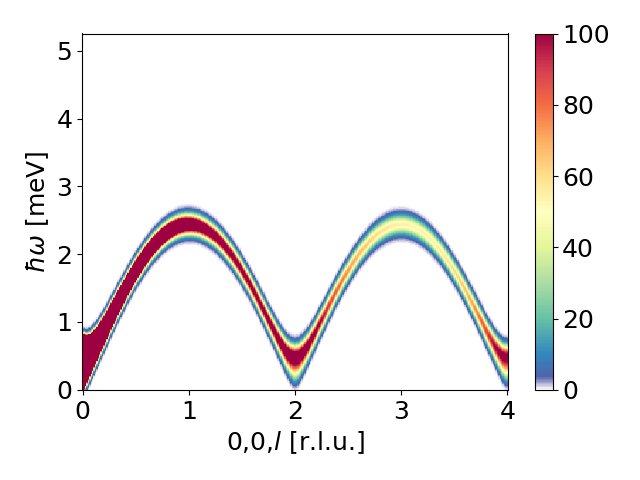}
    \caption{}
    \label{fig:app:spinw00lGamma1}
     \end{subfigure}%
     \begin{subfigure}[b]{0.32\textwidth}
    \includegraphics[width=\linewidth]{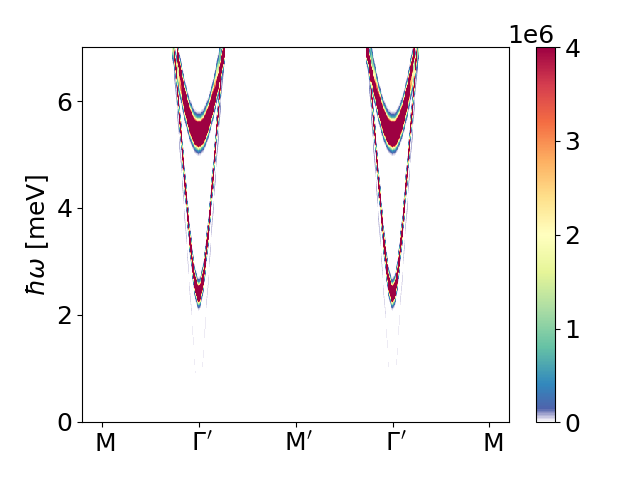}
    \caption{}
    \label{fig:app:spinwH10Gamma1}
     \end{subfigure}

    \begin{subfigure}[b]{0.32\textwidth}
    \includegraphics[width=\linewidth]{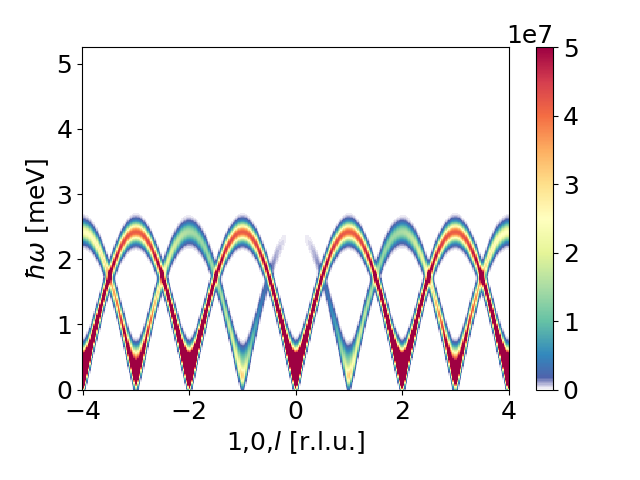}
    \caption{}
    \label{fig:app:spinw10lGamma2}
     \end{subfigure}%
     \begin{subfigure}[b]{0.32\textwidth}
    \includegraphics[width=\linewidth]{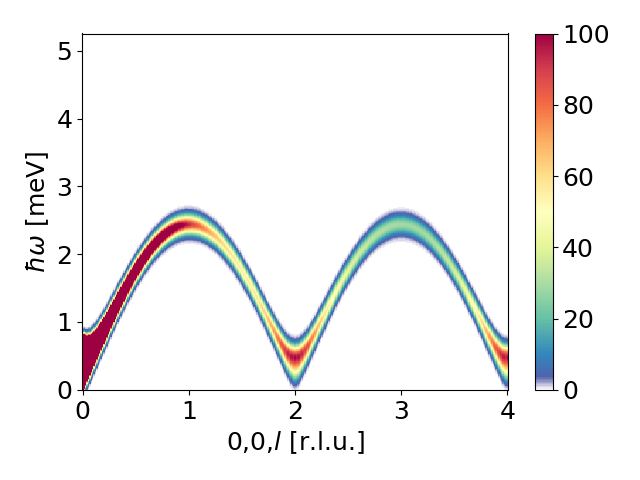}
    \caption{}
    \label{fig:app:spinw00lGamma2}
     \end{subfigure}%
     \begin{subfigure}[b]{0.32\textwidth}
    \includegraphics[width=\linewidth]{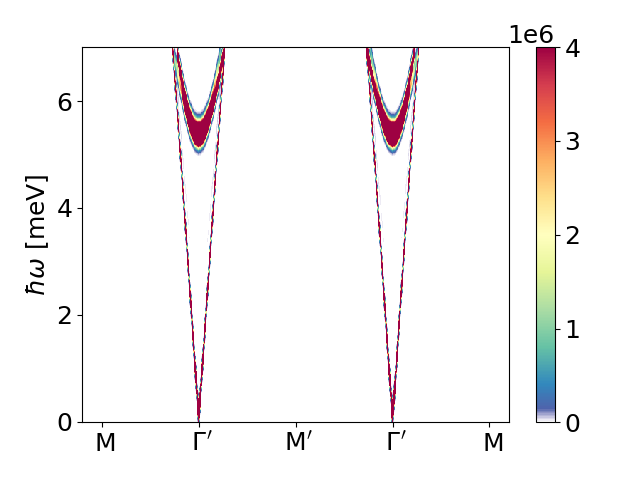}
    \caption{}
    \label{fig:app:spinwH10Gamma2}
     \end{subfigure}

    \begin{subfigure}[b]{0.32\textwidth}
    \includegraphics[width=\linewidth]{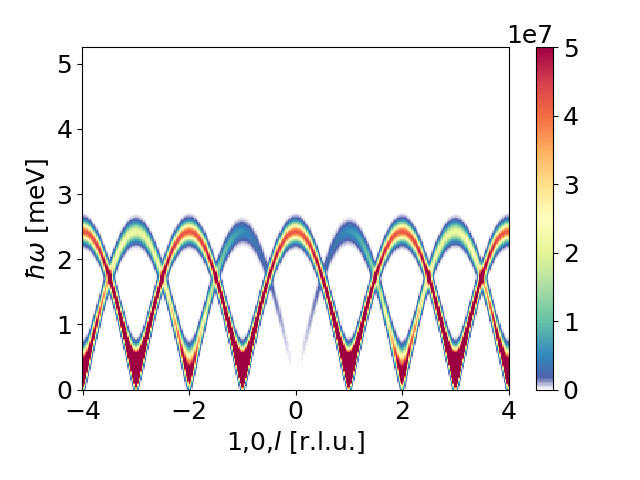}
    \caption{}
    \label{fig:app:spinw10lGamma3}
     \end{subfigure}%
     \begin{subfigure}[b]{0.32\textwidth}
    \includegraphics[width=\linewidth]{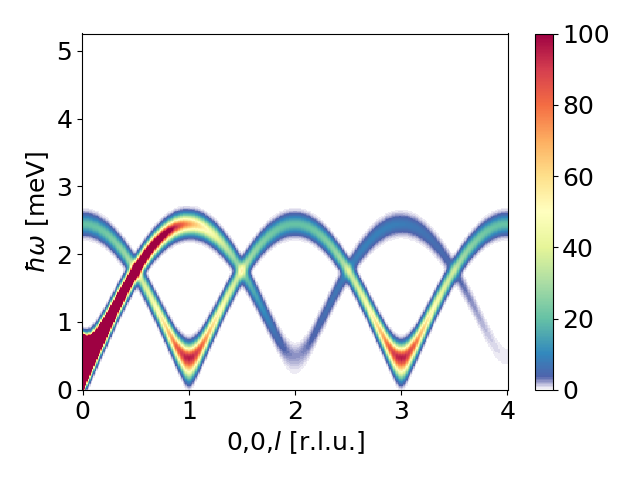}
    \caption{}
    \label{fig:app:spinw00lGamma3}
     \end{subfigure}%
     \begin{subfigure}[b]{0.32\textwidth}
    \includegraphics[width=\linewidth]{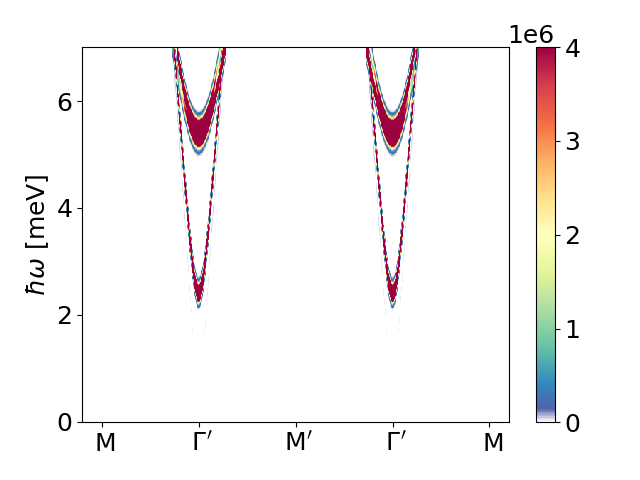}
    \caption{}
    \label{fig:app:spinwH10Gamma3}
     \end{subfigure}

    \begin{subfigure}[b]{0.32\textwidth}
    \includegraphics[width=\linewidth]{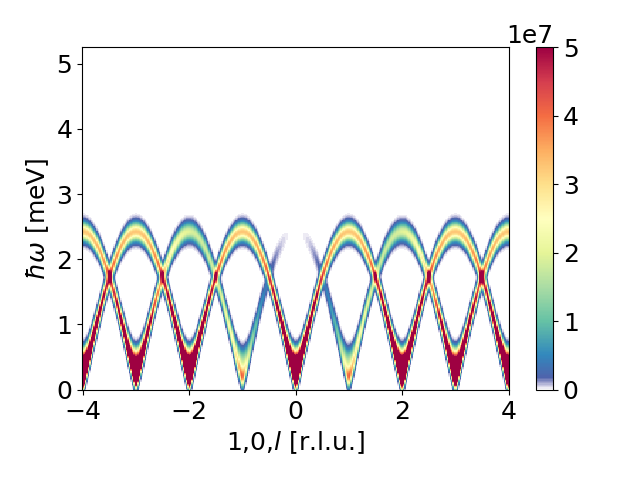}
    \caption{}
    \label{fig:app:spinw10lGamma4}
     \end{subfigure}%
     \begin{subfigure}[b]{0.32\textwidth}
    \includegraphics[width=\linewidth]{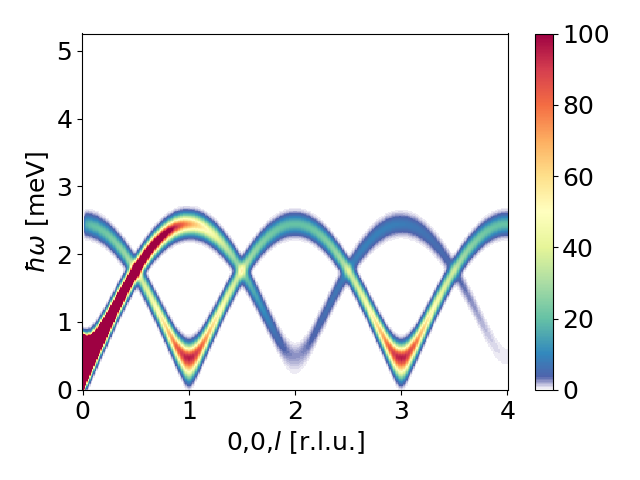}
    \caption{}
    \label{fig:app:spinw00lGamma4}
     \end{subfigure}%
     \begin{subfigure}[b]{0.32\textwidth}
    \includegraphics[width=\linewidth]{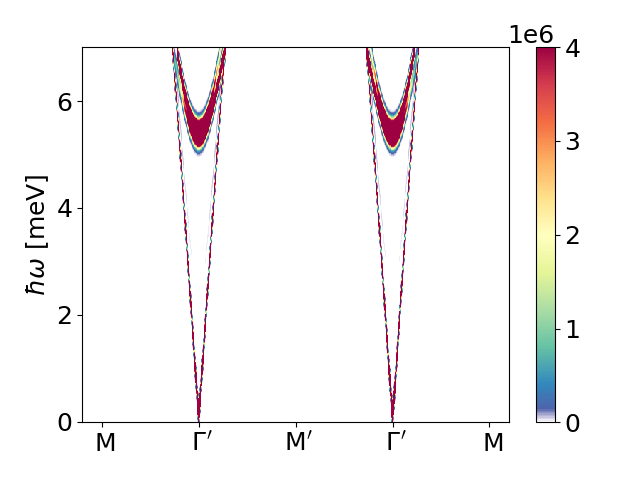}
    \caption{}
    \label{fig:app:spinwH10Gamma4}
     \end{subfigure}
     
    \caption{Model  predictions of the four $\Gamma$ ground states. (a)-(c): $\Gamma_1$ ($P6_3cm$), (d)-(f): $\Gamma_2$ ($P6_3c^\prime m^\prime$), (g)-(i): $\Gamma_3$ ($P6^\prime_3cm^\prime$), (j)-(l): $\Gamma_4$ ($P6^\prime_3c^\prime m$). Colorbars signifies neutron scattering intensity.}
    \label{fig:app:spinWComparison}
\end{figure*}

In order to fully comment on the periodicities and gapped versus non-gapped nature of the spin excitations originating from the four different suggested ground states, SpinW\cite{Toth15} simulations, identical to those presented in Ref.~\onlinecite{Holm2018}, were performed along the three different main directions, plotted in App.Fig.~\ref{fig:app:spinWComparison}. For all of the simulations, the interaction strengths are kept the same as in the previous paper, i.e. $J$ = 2.43~meV, $J_{\rm z1}$ = -0.1533~meV, $J_{\rm z2}$ = 0.1485~meV, and finally $D$ = 0.320~meV and no external magnetic field is imposed. It is noted that the parameters were found for a 2 \% doped sample,\cite{Holm-Dahlin2018Erratum} i.e. Y$_{1-x}$Eu$_x$MnO$_3$ for x = 0.02, but the exact values of the interaction parameters are not of interest in this context.

Looking at the spectra in App.Fig.~\ref{fig:app:spinWComparison}, the main features differ between all of the ground states; the exact amplitudes of the modes change as well as the spin gaps and the spin wave periodicity. Comparing the cut along ($h$, 1, 0) with the data in the ordered phase in Fig.~\ref{fig:Qmap_BZ_p2}~(c) and (d), it is immediately clear that both $\Gamma_2$ and $\Gamma_4$ have spectra in disagreement with the data, and are thus excluded; the same conclusion was reached in Ref.~\onlinecite{Holm-Dahlin2018} as well as when comparing the left hand column with the ordered data in Fig.~\ref{fig:ConstantEnergyH0L}~(a). Only the central column showing the expected scattering intensity along (0,0,$l$) discriminates between $\Gamma_1$ and $\Gamma_3$ where the prior has a periodicity of 2 along $l$ and the latter 1 with a vanishing intensity at $l$ = 2. However, comparing the intensity scale with that of the other two columns, more than 4 orders of magnitude separate the signals. Combining this with the weak signal found in the experiment, Fig.~\ref{fig:QEMaps_0LPhonon}, and the temperature dependency of it, drawing a final conclusion on the ground state structure from the gathered data remains non-viable.

\end{document}